\documentclass[journal]{IEEEtran}
\usepackage{booktabs}
\usepackage{amsfonts}
\usepackage{bm}
\usepackage{amsmath,amsfonts,amsthm,amssymb}
\usepackage{setspace}
\usepackage{Tabbing}
\usepackage{fancyhdr}
\usepackage{lastpage}
\usepackage{extramarks}
\usepackage{chngpage}
\usepackage{algorithmic}
\usepackage{soul,color}
\usepackage{epsfig}
\usepackage{graphicx,float,wrapfig,subfigure}
\usepackage{dsfont}
\usepackage{longtable}
\usepackage{wrapfig}
\usepackage{stfloats}
\usepackage{cite}
\usepackage{graphicx}
\usepackage{array}
\usepackage{multirow}{\tiny }
\usepackage{multicol}
\usepackage{colortbl}
\usepackage{tabularx}
\usepackage{mdwmath}
\usepackage{mdwtab}
\usepackage{color}
\usepackage{verbatim}
\usepackage{amsmath}
\usepackage{tikz}
\usetikzlibrary{calc}
\usepackage{flowchart}
\usetikzlibrary{shapes.geometric, arrows}
\usepackage{overpic}
\usepackage{epstopdf}
\usepackage{url}
\usepackage[colorlinks,linkcolor=black,anchorcolor=black,citecolor=black,urlcolor=black]{hyperref} 

\usepackage{enumitem}
\usepackage[ruled,linesnumbered]{algorithm2e}

\usepackage[justification=centering]{caption} 

\usepackage{siunitx}

\usepackage{color}
\usepackage{hyperref}
\usepackage[hyphenbreaks]{breakurl}
\makeatletter %
\let\myorg@bibitem\bibitem
\def\bibitem#1#2\par{%
	\@ifundefined{bibitem@#1}{%
		\myorg@bibitem{#1}#2\par
	}{%
		\begingroup
		\color{\csname bibitem@#1\endcsname}%
		\myorg@bibitem{#1}#2\par
		\endgroup
	}%
}

\makeatother %

\allowdisplaybreaks

\usepackage{footnote}
\makesavenoteenv{tabular}
\makesavenoteenv{table}

\begin{document}

\title{{
        Safe Reinforcement Learning for Power System Control: A Review
        }}

\author{
Peipei~Yu,~\IEEEmembership{Student Member,~IEEE,}
Zhenyi~Wang,~\IEEEmembership{Student Member,~IEEE,}
Hongcai~Zhang,~\IEEEmembership{Senior Member,~IEEE,}
and~Yonghua~Song,~\IEEEmembership{Fellow,~IEEE}

\thanks{This paper is funded in part by the Science and Technology Development Fund of Macau SAR (File no. 001/2024/SKL, and File no. 0053/2022/AMJ). (Corresponding author: \textit{Hongcai Zhang}.)}

}

\maketitle

\begin{abstract}
The large-scale integration of intermittent renewable energy resources introduces increased uncertainty and volatility to the supply side of power systems, thereby complicating system operation and control. Recently, data-driven approaches, particularly reinforcement learning (RL), have shown significant promise in addressing complex control challenges in power systems, because RL can learn from interactive feedback without needing prior knowledge of the system model. However, the training process of model-free RL methods relies heavily on random decisions for exploration, which may result in ``bad" decisions that violate critical safety constraints and lead to catastrophic control outcomes. Due to the inability of RL methods to theoretically ensure decision safety in power systems, directly deploying traditional RL algorithms in the real world is deemed unacceptable. Consequently, the safety issue in RL applications, known as safe RL, has garnered considerable attention in recent years, leading to numerous important developments.
This paper provides a comprehensive review of the state-of-the-art safe RL techniques and discusses how these techniques can be applied to power system control problems such as frequency regulation, voltage control, and energy management. We then present discussions on key challenges and future research directions, related to convergence and optimality, training efficiency, universality, and real-world deployment.
\end{abstract}

\begin{IEEEkeywords}
Safe reinforcement learning, frequency regulation, voltage control, energy management, power systems.
\end{IEEEkeywords}

\section{Introduction}
With the intensifying change in global climate, promoting carbon neutrality has become a consensus worldwide. As a critical energy infrastructure, it is necessary to drive energy transition in power systems to reduce carbon emissions\cite{lopion2018review}. Energy transition mainly facilitates two changes to traditional power systems:
1) The system energy structure changes, where renewables (e.g., solar photovoltaic and wind power) will probably become the main power source with its increasing penetration. This change leads to greater uncertainty and volatility on the power supply side, further complicating the real-time match with the dynamic power demand side. 
2) With the integration of distributed energy resources (e.g., energy storage systems and electric vehicles), the traditional centralized and large-scale control is undergoing a shift to a distributed and collaborative one. Moreover, complex system characteristics of massive resources make it hard to model the network, such as unknown model parameters.
Therefore, to cope with the above uncertainty, volatility, distribution, and complexity, power systems are undergoing a transformation to smart grids, becoming more flexible and intelligent\cite{borlase2017smart}. Specifically, unpredictable fluctuations and distributed model complexity brought by this transformation require real-time monitoring and strategic control for power systems, based on advanced information and communication technology and intelligent control methods.

In particular, reinforcement learning (RL) has been considered a prominent approach to overcome these challenges in smart grids. On the one hand, RL can learn from interaction feedback without prior knowledge of the system model\cite{cao2020reinforcement}. On the other hand, RL can utilize neural networks to establish data-driven models for uncertain environment descriptions, enabling well-trained agents to adapt to environment changes continually. Over the past decade, RL has achieved great success in complex control problems of power systems, e.g., frequency regulation and voltage control\cite{li2023deep, Lina_RL_review}.
Generally, before deploying RL controllers on real-world systems, existing RL-related studies in power systems can only train RL controllers on high-fidelity simulations while not directly on real-world systems. This is because traditional RL training cannot theoretically guarantee decision safety in a real-world system\footnote{Traditional RL training process relies on random decisions for explorations, which means the RL agent probably makes ``bad" decisions that violate critical safety constraints and lead to catastrophic control results}. 
However, considering the system gap between simulations and the real world, controllers that are trained completely based on simulations cannot ensure effectiveness in real-world systems. Hence, it is necessary to achieve safe training on real-world systems to bridge the above gap from theory to practice, which has triggered research on the safe RL\cite{pecka2014safe}.

Safe RL is considered a sub-field within RL that is envisioned to compensate for the limitations of traditional RL in safety issues, which was first defined by J. Garcıa in 2015\cite{2015_review_safe_RL}. Recently, safe RL has attracted surging attention in power system control. In 2020, authors in \cite{li2019constrained} first applied safe RL techniques in power systems, for electric vehicle (EV) charging scheduling.  Although safe RL applications have been mentioned in a few reviews (e.g., energy system\cite{wei2022incorporating}), to our knowledge, this is the first paper to provide a review of safe RL techniques applied in power systems. 
We first summarize various state-of-the-art safe RL techniques, then exemplify how these techniques are applied to control problems in power systems, and finally discuss the key challenges and perspectives. Overall, the key contributions of this work are threefold:
1) We present a comprehensive and structural overview of safe RL techniques, in terms of basic concepts and theoretical fundamentals, summarizing two technical categories from whether the safe module is coupled with the traditional RL framework.
2) We present the effective application of safe RL techniques in modeling, safe module design, and implementation, by selecting three key applications in power systems, i.e., frequency regulation, voltage control, and energy management.
3) We discuss the key challenges and perspectives for applying safe RL in power systems regarding convergence and optimality, training efficiency, universality, and real-world deployment.

The rest of this paper is organized as follows: Section \ref{sec:2} introduces the basic concept of RL, and describes two categories of the state-of-the-art safe RL techniques. Section \ref{sec:3} provides a comprehensive review of RL applications to three critical power system problems, i.e., frequency regulation, voltage control, and energy management, illustrating typical mathematical models. Section \ref{cases} discusses the key challenges (e.g., optimality, efficiency, and universality) and potential future directions of safe RL techniques. Finally, Section \ref{conclude} presents our conclusions.
\begin{figure*}
    \centering
    \includegraphics[width=2.1\columnwidth]{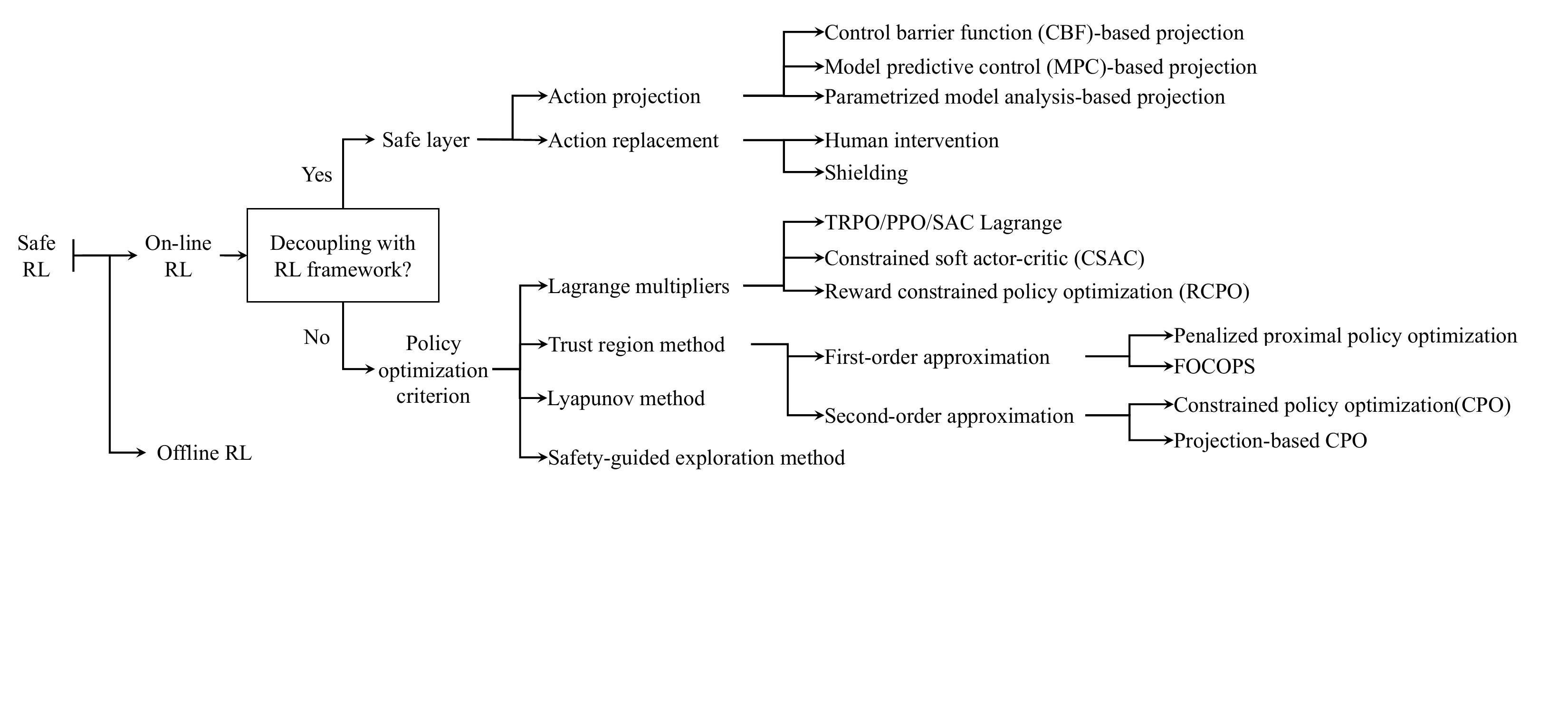}
    \vspace{-28mm}
    \caption{The structure of safe RL methodology.}
    \label{fig_safe_RL}
\end{figure*}

\section{Methodology of safe reinforcement learning}
\label{sec:2}
This section first introduces the basic formulation and key variables of the traditional RL algorithm in subsection \ref{sec:2.1}, which is the preliminary of safe RL. Then, two categories of state-of-the-art safe techniques are introduced in detail, which consider the safety in online RL through a safe layer (in subsection \ref{sec:2.2}) and transforming policy optimization criterion (in subsection \ref{sec:2.3}). For each category, we summarize the key ideas of different techniques and analyze their applicable scenarios. Fig. \ref{fig_safe_RL} shows the overall classification of safe RL techniques summarized in this paper, along with relevant literature references. 
It is noted that, offline RL learns optimal strategy from a static dataset, which avoids interactions with the physical environment. Although avoiding direct interactions can guarantee constraint safety during the RL training process, the problem of distribution shift between the static dataset and the real-world environment also brings challenges to strategy solving\cite{offlineRL_review_1}. At present, the convergence and optimality of offline RL algorithms lack theoretical guarantees. The majority of research in offline RL focuses on addressing the problem of distribution shift, rather than on constraint safety considerations\cite{lesage2022batch, offlineRL_review_2}. Hence, as a special category of safe RL, this paper does not delve into a detailed discussion on offline RL.

\subsection{Basic formulation of reinforcement learning}
\label{sec:2.1}
RL is a branch of machine learning that focuses on training a controller to find optimal sequential decisions in an uncertain environment. As a typical modeling technique, the Markov Decision Process (MDP) is often employed to describe sequential decision-making problems. The mathematical framework provided in an MDP includes four essential components to describe the interaction between the \textbf{environment} and \textbf{agent}, which are the \textbf{state} space $s\in\mathcal{S}$, the \textbf{action} space $a\in\mathcal{A}$, the \textbf{transition} function $\mathbb{P}(\cdot|s,a):\mathcal{S}\times\mathcal{A}\rightarrow P(\mathcal{S})$, and the \textbf{reward} function $r:\mathcal{S}\times\mathcal{A}\rightarrow\mathbb{R}$, respectively. Here, $P(\mathcal{S})$ denotes a distribution on the state space. Hence, in power systems, a specific control problem is first described as an MDP mathematically, by defining the four aforementioned components. Then, the MDP can be solved by different RL algorithms.

\begin{figure}
    \hspace{1em}
    \includegraphics[width=0.9\columnwidth]{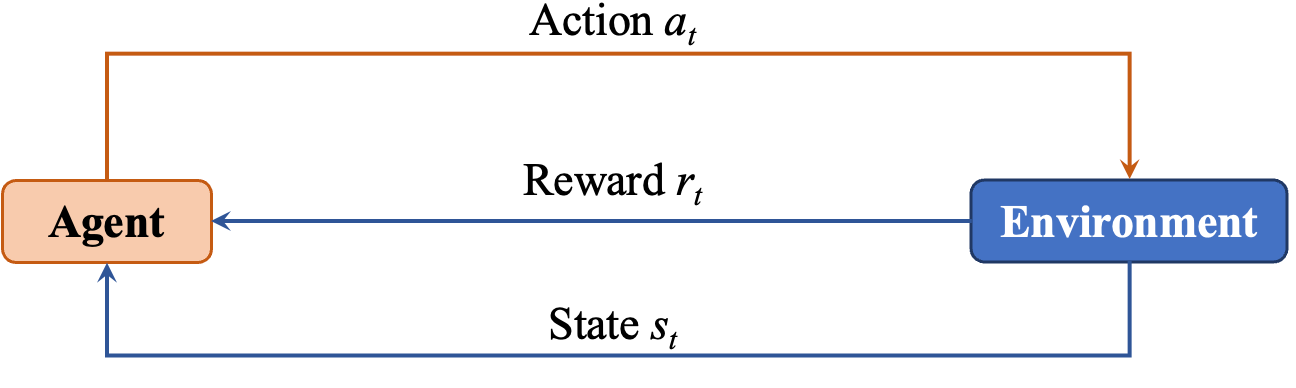}
    \vspace{-0mm}
    \caption{Illustration of a Markov Decision Process.}
    \label{fig_MDP}
\end{figure}

To show the continuous interaction in an MDP, Fig. \ref{fig_MDP} displays the relationship between the four components $<\mathcal{S},\mathcal{A},\mathbb{P},r>$. For each time step $t=\{0,1,...\}$, the environment captures the current system operation state $s_t$ through defined observed variables, and sends it to the agent. Then, based on the received state information $s_t$, the agent makes an action decision $a_t$ and executes it. Further, the environment returns the reward $r_t$ as feedback for the action. Finally, the environment goes to the next state $s_{t+1}$, following the transition probability $\mathbb{P}(s_{t+1}|s_t,a_t)$. The agent's rule to take what action given a certain state $s_t$ is called \textbf{policy} $\pi(a|s):\mathcal{S}\rightarrow P(\mathcal{A})$, mapping from the state space to a distribution on the action space.
For the agent in an MDP, the optimal policy $\pi^*$ means the policy can help the agent receive the maximum expected cumulative reward $J_R^\pi$:
\begin{align}
    &\pi^*: \arg\max_{\pi} J_R^\pi
    =\mathbb{E}_{\tau\sim\pi}[\sum\nolimits_{t=0}^\infty \gamma^t r(s_t,a_t)]
    ,\label{eqn_RL_obj}
\end{align}
where $\tau=\{s_0,a_0,s_1,a_1,...\}$ is a state-action trajectory and $\tau\sim\pi$ means that the action in $\tau$ is selected based on the policy $\pi(\cdot|s_t)$. Parameter $\gamma\in[0,1)$ is a discount factor that considers the reward from future steps. Traditional RL techniques for solving Eq. (\ref{eqn_RL_obj}) have been summarised in \cite{Lina_RL_review}.

To solve the optimal policy of MDPs, the model-free RL algorithm is the most popular in power systems, because model-based RL algorithms and dynamic programming rely heavily on perfect environmental assumptions (e.g., accurate state transition probability $\mathbb{P}$) and high-complexity computation\cite{RL_book}. 
Generally, in the initial policy learning stage, model-free RL algorithms introduce a random process for action explorations to collect adequate experiences. However, such conventional RL methods are only suitable for inherently safe systems or simulators, allowing agents to engage in unconstrained ``trial-and-error". This is because in real-world power grids, random action explorations could lead to extremely dangerous situations, or even result in significant safety incidents \cite{srinivasan2020learning}. Therefore, the safe concept is introduced into RL, forming safe RL, to address the security considerations of RL applications in real-world power systems.

\subsection{Safe reinforcement learning by adding safe layer}
\label{sec:2.2}
For the RL application in power systems, the agent's action following the learned policy is \textit{safe} if the system's state satisfies its operation constraints after action execution, especially for hard constraints\cite{2015_review_safe_RL}. However, during the training process, the agent learns more about the environment only through explorations, where most exploration techniques in RL are blind to the risk of actions (e.g., heuristics or $\varepsilon -$greedy)\cite{yang2021exploration}. Exploration is essential for the agent's training, so the random perturbation to actions is hard to avoid in RL. Therefore, a straightforward idea is to design a \textbf{safe layer} before executing every action. The safe layer is expected to verify whether the action is safe for all constraints, and tune unsafe actions to safe ones. Based on the verified result, the safe layer can modify the action through different techniques to ensure the final executed action is safe. 

\begin{figure}
    \hspace{1em}
    \includegraphics[width=0.9\columnwidth]{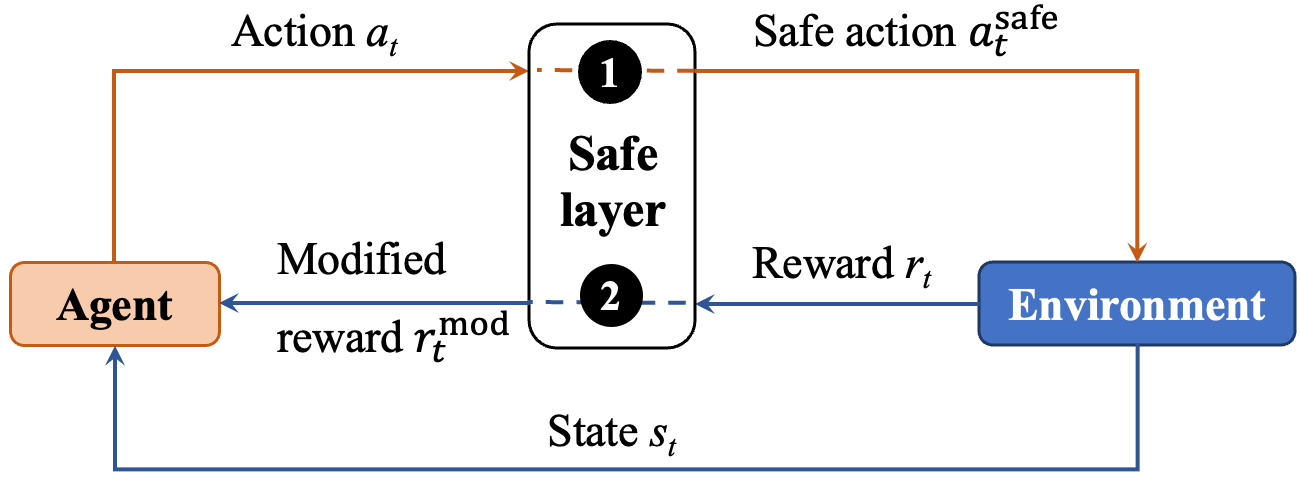}
    \caption{Combination scheme of the safe layer and RL. At (1), the safe layer intercepts unsafe actions to become safe ones. At (2), the safe layer modifies the reward based on the interception degree, returning effective feedback.}
    \label{fig_safe_layer_scheme}
    \vspace{-2mm}
\end{figure}

Adding a safe layer is a main tendency in safe RL, because this is a universal technique decoupling with the traditional RL framework. That is, the introduced safe layer can combine with various RL algorithms, whether based on the actor-critic scheme\cite{grondman2012survey}, policy gradient\cite{DPG}, or policy optimization\cite{TRPO}. Fig. \ref{fig_safe_layer_scheme} displays the combination scheme of the safe layer and the RL agent, which involves two key steps: 1) At each time step, the safe layer intercepts unsafe actions $a_t$ to become safe ones $a^\text{safe}_t$; 2) As the agent's feedback, the safe layer modifies the original reward to reflect the penalty of the interception degree, from $r_t$ to $r_t^\text{mod}$. Based on the safe layer scheme, many researchers have proposed different design techniques to tune actions, including two main categories: \textit{action replacement} and \textit{action projection}\cite{krasowski2022provably}, which are introduced in detail as follows. 

\subsubsection{Action replacement}\label{sec_action_replace}
Fig. \ref{fig_safe_layer_scheme}(a) shows the concept of action replacement, where the key technique is how to obtain safe actions $a^\text{safe}_t$ in a safe action space $\mathcal{A}^\text{safe}$ to replace the original unsafe ones $a_t$. Recently, researchers have proposed to obtain a single safe action via human feedback \cite{gu2023human} or from a shielding/blocked mechanism\cite{shielding_safe_layer, fulton2018safe} (e.g., a failsafe planner \cite{9811698}).

\textit{1) Human intervention:}
For safety-critical scenarios, human expert experiences can ensure great safety to avoid unsafe actions made by RL agents\cite{human_intervention_safe_layer}. Hence, leveraging human expertise to guide the exploration of agents is a natural idea to enhance safety, where the most common methods include the \textit{interruption mechanism} and \textit{expert guidance}.

The interruption mechanism aims to interrupt the final executed action directly when it is considered dangerous by drawing on human experiences, and then replace with a safe action. This method can cope with ``catastrophic actions" that the human overseer deems unacceptable under any circumstances, which only relies on human experts but requires no model information.
However, RL training iteration is up to millions; it is not practical for a human expert to constantly supervise an RL agent for a million timesteps. For this consideration, achieving automating oversight using human expert experiences has been paid attention in recent years. Saunders et al.\cite{human_intervention_safe_layer} proposed to train a ``blocker" to mimic human interruption, which includes two steps: 1) manual supervision stage to collect binary label of ``whether the manual interruption is implemented"; 2) ``blocker" training stage to mimic the human interruption. It should be noted that the manual supervision stage cannot stop until the "blocker" performs well on the testing dataset.
However, the limitation is that the ``blocker" can only handle relatively simple accidents. When facing more complex environments, the ``blocker" takes more than one year to implement the manual supervision stage, with high time costs. To solve the above issue, Prakash et al.\cite{prakash2019improving} proposed a hybrid safe RL scheme for reducing the time cost, by combining a model-based prediction module with the ``blocker". In addition to training a ``blocker" to identify dangerous actions, humans prefer to stop the action immediately when confronted with potential danger. Inspired by this, Sun et al. \cite{sun2021safe} and Eysenbach et al.\cite{eysenbach2017leave} both developed a safe RL framework for early response to potential dangers, where the former introduced a solution to early terminated MDP (ET-MDP) and the latter proposed to automatically reset the environment.

The expert guidance introduces \textit{curriculum learning} \cite{narvekar2020curriculum} into the conventional RL framework to guarantee safety, whose main idea is to imitate the learning process of humans, from simple tasks to difficult tasks\cite{wang2021survey}, speeding up the training efficiency. In 2020, Turchetta et al.\cite{turchetta2020safe} adopted curriculum learning to ensure safety in RL for the first time, where an agent (i.e., student) learns from the automatic guidance of a supervisor (i.e., teacher). The supervisor needs to automatically design the course for the agent, according to the agent's learning progress and behavioral data distribution. Hence, the safe exploration of the agent completely depends on whether its supervisor is well-trained. However, the challenge in curriculum learning is how to train the supervisor with limited samples, when designing complex learning tasks. To cope with this issue, Peng et al.\cite{peng2022safe} proposed an expert-guided policy optimization, which combines offline RL to stabilize the training of the supervisor through the off-policy partial demonstration. Nonetheless, high-cost offline human intervention increases the reliance on experts. Then, in 2022, Li et al.\cite{li2022efficient} developed a novel human-in-the-loop learning by designing a special mechanism to mitigate the delayed feedback error, which can effectively reduce the reliance on experts over time and improve the supervisor’s autonomy. With this self-learning method, one may cause a quite conservative policy. 

In summary, human intervention requires experts to help the agent become self-learning automatically. Although this approach can be both safe during training and deployment, the high cost of human intervention should be taken into account in real-world applications. 

\textit{2) Shielding:}
The concept of ``shielding" in RL was first proposed by Alshiekh et al.\cite{shielding_safe_layer} in 2018, to ensure constraint safety during RL training. The key idea is that, the shielding process will be triggered when the output action is unsafe, and then an alternate safe action is used to override the original one. Hence, the implementation of the shielding mainly involves two significant works: 1) the design of the shielding trigger, and 2) the design of backup (safe) policies. 

The design of the shielding trigger is hard to design. Generally, the shielding method is more suitable for scenarios when safety conditions and constraints can be clearly defined (e.g., no speeding), because the shielding trigger would be easier to design. For complex scenarios with constraints that are hard to define, some researchers have proposed \textit{model predictive shielding} (MPS) to handle deterministic closed-loop environment dynamics\cite{bastani2021safe, zhang2019mamps} or stochastic environment dynamics\cite{li2020robust, wang2023enforcing}. This promising MPS method can perform shielding on-the-fly instead of ahead-of-time, by checking whether a single state is safe in real-time\cite{MPC_linear_2018}. For common deterministic shielding methods, there are only two system states, either safe or unsafe. However, the same action may lead to different state results following ambient uncertainties. Hence, probabilistic shielding is further proposed to cope with uncertainties\cite{jansen2020safe} through formal verification to compute the probabilities of critical decisions. 

The design of backup policies is also one of the research focuses in shielding, where MPC-based backup controllers are the most common choice. Li and Bastani \cite{li2020robust} used a robust nonlinear MPC to compute a backup policy in stochastic environment dynamics. To improve the safety of the backup policy, Bastani \cite{bastani2021safe} further defined the backup policy with two choices: an invariant and a recovery policy. The invariant strategy can keep the agent moving around the safe equilibrium point, and the recovery strategy can move the agent to the safe equilibrium point. The key idea is that the controller can determine which backup policy to use based on the distance between the agent and the safe equilibrium point.

In addition, as step 2 shown in Fig. \ref{fig_safe_layer_scheme}, the reward after shielding $r^\text{mod}$ needs to provide feedback of shielding interception for agents. One usually has two different designs: 1) assigning a large punishment to learn that selecting $a_t$ at state $s_t$ is unsafe, $r^\text{mod} \textless r_t$; 2) remaining the same with original reward, $r^\text{mod}=r_t$. For the former approach, the agent can learn from the punishment feedback, so the shield is no longer needed in the execution phases for well-trained agents. For the latter, the agent cannot learn to avoid unsafe actions, which are always corrected to safe ones by the shield without feedback. Hence, the shield is still needed in the execution phases even for well-trained agents.

Therefore, compared with human intervention-based methods, shielding is an automated security mechanism with low costs, which dynamically adjusts the decision space through predefined rules or real-time calculated risk assessments. However, the limitation of shielding is the low adaptability in dynamic and complex environments, which is a model-based technique. For complex tasks, it is difficult to provide sufficient prior knowledge to build comprehensive shielding from all dangers\cite{carr2023safe}, as human experts do.

\subsubsection{Action projection}\label{sec_action_projection}
\begin{figure*}
    \hspace{1cm}
    \includegraphics[width=2.05\columnwidth]{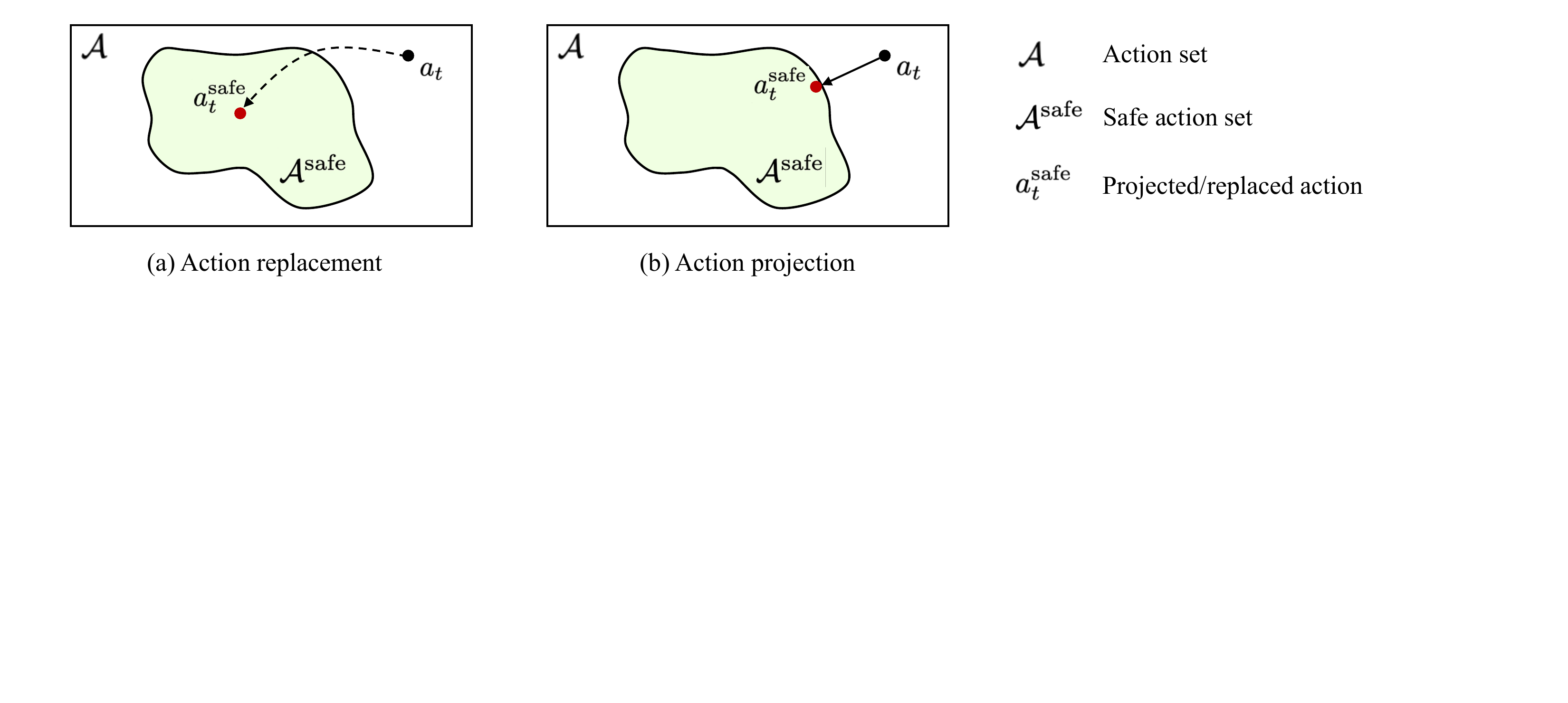}
    \vspace{-50mm}
    \caption{Concept and differences between the action projection and action replacement. Here, action replacement (a) replaces unsafe actions with self-defined actions from the safe action space. Action projection (b) projects the agent's unsafe actions to the closest action in the safe action space.}
    \label{fig_replace_projection}
\end{figure*}

After reviewing the research work that applies safe RL to power systems, we find out that action projection is the most popular method to deal with constraint safety issues \cite{wang2023secure,yu2023district,zhao2023barrier,rokhforoz2023safe}. As shown in Fig. \ref{fig_replace_projection}(b), action projection aims to project the original unsafe action to an action in the safe space $\mathcal{A}^\text{safe}$ that is closest to itself, where the projection rule design relies on model-based optimization programming.
In a theoretical way, the common design of projection rules can be categorized broadly based on three techniques: CBF, MPC, and parametrized model analysis (as summarized in Fig. \ref{fig_safe_RL}). Among the above three designs, the similarity is that, they all need to obtain the closest safe action $a^\text{safe}_t$ by solving an optimization problem at every time step. The difference is how to define the objective and constraints of the optimization problem, and what the system model assumptions of physical environments are. Three projection rule designs are introduced in detail as follows. 

\textit{1) Control barrier function:}
The basic idea of CBF is to define a safe region (so-called ``safe set") by creating a ``barrier", which can effectively prevent the agent from stepping into the unsafe region and staying inside the safe set. This technique can effectively deal with hard constraints that must be satisfied, because the safe set defined by CBF possesses forward invariance property \cite{yang2020safe}. 
Specifically, a safe set $\mathcal{C}$ is required to be defined by the super-level set of a continuously differentiable function $h:\mathbb{R}^n\rightarrow\mathbb{R}$, by $\mathcal{C}:\left\{s_t\in\mathbb{R}^n:h(s_t)\geq 0\right\}$. To maintain the constraint safety during every training step, the RL agent can only be limited to learning and exploring within the safe set $\mathcal{C}$. Here, one challenge is how to design functions $h$ in CBF that can guarantee the safe set $C$ possesses forward invariant property. 
For instance, Cheng et al. \cite{CBF_safe_layer} proposed an \textit{affine} CBF based on discrete-time formulations. Emam et al. \cite{RCBF_safe_layer} further extended the design of $h$ from an affine form to a finite union form of convex hulls, which can effectively capture non-convex disturbances in environment systems.

Here, we simply give an example in an affine environment system to show the corresponding design of safe sets, which also supports a more general form. If there exists $\eta\in[0,1]$ satisfying the condition: $\sup_{a_t\in\mathcal{A}}[h(s_{t+1})+(\eta-1)h(s_t)]\geq 0$, the differentiable function $h$ is defined as a discrete-time CBF. Then, the proposed safe set can compensate for the unsafe action by $\Delta a_t$, based on the CBF controller $\pi^\text{CBF}$ as:
\begin{align}
    &a_t^\text{safe} = a_t+ \pi^\text{CBF}(s_t,a_t)=a_t + \Delta a_t,
    \label{CBF_compensate}
\end{align}
where $\Delta a_t$ represents the compensation value at time step $t$. Further, in the safe layer, the objective of the projection is designed to provide minimal control intervention to original actions $a_t$. The constraints consist of the upper/lower limits of final actions and the predefined safe set. Finally, at each time step, the model-based optimization programming for the action projection is formulated as:
\begin{subequations}
\label{CBF_main}
    \begin{align}
        (\Delta a_t,\varepsilon)=
        &\mathop{\arg\min}\limits_{\Delta a_t,\varepsilon}\quad \Vert \Delta a_t \Vert_2+K_\varepsilon \varepsilon,\label{QP_objective}\\
        \text{s.t.:}~
        &h(s_{t+1})\geq (1-\eta)h(s_t)-\varepsilon,\label{CBF_constraint}\\
        &a_t^\text{safe} = a_t + \Delta a_t,\\
        &\underline{a}\leq a_t^\text{safe}\leq \overline{a},
        \label{CBF_cons_2}
    \end{align}
\end{subequations}
where $\varepsilon$ is a slack variable for the safe set, and $K_\varepsilon$ is a large constant (e.g., 10$^{12}$) for penalization when safety cannot be enforced. Note that, different CBF designs will derive different constraint formulations in Eq. (\ref{CBF_constraint}), where the safety of CBF needs to be reproved theoretically\cite{CBF_good_1, CBF_safe_layer, RCBF_safe_layer}. Although CBF can guarantee safety for an infinite time, finding a suitable and well-designed CBF is not easy in practice. 

Conventionally, CBF is a model-based method that requires the model of system dynamics (e.g., in Eq. (\ref{CBF_constraint}), variable $h(s_{t+1})$ should be expressed). Hence, without prior knowledge of the system model, integrating the model-based CBF with a model-free RL framework is another challenge. Currently, researchers have proposed several model estimation methods to compensate for the unknown system dynamics, such as the Gaussian process\cite{wang2018safe, CBF_safe_layer}, the iterative search algorithm\cite{wang2018permissive}, sparse optimization \cite{ohnishi2019barrier}, and state transformation\cite{yang2019safety}. The model estimation accuracy of system dynamics is also a significant factor for the constraint safety effects, where a high-accuracy model brings a high-probability safety guarantee. However, measurement errors and environmental noise are hard to be considered perfectly in constraint models.

\textit{2) Model predictive control:} MPC is considered a common methodology for constrained control, which can exploit the data reliably to take safety constraints into account\cite{MPC_review}. Recently, researchers have proved that the combination of the MPC and RL algorithms can achieve a safe and high-performance system operation. 
Compared with the CBF technique, MPC-based safe RL also designs the action projection by a model-based optimization programming. However, the MPC technique guarantees constraint safety through a predictive safety filter, but not through defined safe sets. Hence, unlike the required forward invariant property in CBF, an MPC-based learning process has better universality without design requirements. 
Specifically, based on the MPC technique, at each time step $t$, the projected safe action for original action $a_t$ is solved by the following optimization problem:
\begin{subequations}
\label{MPC_main}
    \begin{align}
        \min_{a^\text{safe}_{j|t},s_{j,t}}
        &||a_t-a^\text{safe}_{0,t} ||,\label{MPC_objective}\\
        \text{s.t.:}~
        &s_{0,t} = s_t,\\
        &s_{N,t} \in \mathcal{S^\text{safe}},\\
        &s_{j+1,t} = f(a_{j,t},s_{j,t}),\quad \forall j\in\mathcal{J}_{[0,N-1]},\label{MPC_model_cons}\\
        &a_{j,t} \in \mathcal{A}, \quad \forall j\in\mathcal{J}_{[0,N-1]}\\
        &s_{j,t} \in \mathcal{S},  \quad \forall j\in\mathcal{J}_{[0,N-1]}
    \end{align}
\end{subequations}
where $s_{j,t}$ is the $j$ time steps-ahead state prediction, computed at time $t$ (i.e., $s_{0,t}=s_t$); $\mathcal{S^\text{safe}}$ is a terminal safe set; $f(a,s)$ is the system dynamic model; $N$ is the prediction horizon. 
The problem (\ref{MPC_main}) solves an $N$-step input sequence $\{a^\text{safe}_{j,t}\}$ that drives the system to the terminal safe set, guaranteeing safety for all future time steps (detailed proof in \cite{MPC_RL_refer}). Then, the solved first input $a^\text{safe}_{0,t}$ is selected as the safe action at time $t$, which is expected to be as ``close” as possible to the agent's original output. 

However, MPC-based safe RL methods rely heavily on the system model $f(a,s)$ shown in Eq. (\ref{MPC_model_cons}), because the model accuracy directly influences the prediction result of future $N$-steps state safety. Hence, one drawback of MPC is that it is usually not robustly safe, because it cannot encode uncertainties (e.g., environment noises and measurement errors) into the optimization problems, especially for nonlinear systems\cite{MPC_review}. Currently, some works have focused on improving the robustness of MPC-based RL in safety guarantees, including linear\cite{MPC_linear_2018} and non-linear systems\cite{MPC_non_linear_2021predictive}. Although virtually any RL controller can be enhanced with safety guarantees using MPC, the resulting performance of the overall system remains to be investigated.

\textit{3) Parameterized model analysis:}
Theoretically, for constraint safety, the most reliable method is to design an action projection rule through the analysis of known parameterized models. This is because the parameterized model-based optimization can predict the future operation state of systems more accurately to judge the safety\cite{RCBF_safe_layer}. However, the system model assumption becomes quite strong for most practical scenarios, requiring an accurate parameterized model of system operation constraints.

Recently, in power systems, some researchers have tried to apply parameterized model analysis forming an action projection in the safe layer, including solving the optimal power flow problem\cite{yi2023real}, the voltage control problem\cite{kou2020safe}, demand-side resource management\cite{yu2023district,guo2023safe}, and electricity market bidding\cite{rokhforoz2023safe}. Although systems constraint models in the above-mentioned scenarios vary, their proposed objectives for the optimization programming are the same, which aim to minimize the distance between corrective safe actions $a^\text{safe}_t$ and original unsafe actions $a_t$.
While an advantage is that the parameterized model-based optimization can effectively handle hard system constraints, the drawbacks are: 1) the projection design in safe layers relies heavily on the parameterized system model, which is not universal and cannot directly extend to different scenarios; 2) the system uncertainty is hard to be parameterized and considered into the optimization problem, especially involving stochastic human behaviors.

Therefore, the key problem in action projection is formulating the constraint based on different system conditions. Specifically, the CBF-based method requires little system model knowledge, while putting a high requirement on the barrier function design, such as Lipschitz condition. The effectiveness of MPC-based methods depends on the accuracy of the applied model, where the robustness in non-linear systems remains to be investigated. The parameterized model-based method is more reliable for constraint safety, while the model assumption is quite strong in practice. However, when a system is a black box to its controller, the optimization problem cannot be formulated to apply the action projection technique.

\subsection{Safe reinforcement learning by transforming policy optimization criterion}
\label{sec:2.3}
The above safe layer-based technique adopts an extra safe layer that is decoupled with RL algorithms. This subsection presents another safe RL technique considering constraint safety by changing the agent's policy optimization criteria, which is coupled with RL algorithms. As shown in Eq. (\ref{eqn_RL_obj}), the goal of conventional RL is to maximize cumulative rewards, ignoring the damage that constraint violations cause to the agent. That is, the objective function in MDP formulation lacks the description of constraint violation risks or losses. Hence, to describe the constraint violation mathematically in safe RL problems, the traditional formulation of MDP in RL is reformulated into the following CMDP. 

\subsubsection{Extended formulation of safe reinforcement learning}
A CMDP extends the MDP framework $<\mathcal{S}, \mathcal{A}, \mathbb{P}, r>$ by introducing constraints to restrict the allowable policies\cite{altman2021constrained}. Specifically, the MDP is augmented with an auxiliary cost function $C$ and the corresponding threshold $d$, where the cost function $C:\mathcal{S}\times\mathcal{A}\rightarrow\mathbb{R}$ maps station-action pair to violation costs. Similar to the expected cumulative reward $J_R^\pi$ in Eq. (\ref{eqn_RL_obj}), the expectation over the violation cost $J_{C_i}^\pi$ is denoted by:
\begin{align}
    &J_{C}^\pi
    =\mathbb{E}_{\tau\sim\pi}[\sum\nolimits_{t=0}^\infty \gamma^t C(s_t,a_t)]
    .\label{eqn_CMDP_cost}
\end{align}
Thus, the reformulated policy optimization problem for a CMDP becomes:
\begin{subequations}
\label{CMDP_main}
    \begin{align}
        \pi^*: \arg\max_{\pi} 
        &J_R^\pi 
        =\mathbb{E}_{\tau\sim\pi}[\sum\nolimits_{t=0}^\infty \gamma^t r(s_t,a_t)] \\
        \text{s.t.:}~
        &J_{C}^\pi
        \leq d
        .\label{eqn_CMDP_constraint}
\end{align}
\end{subequations}

As discussed in subsection \ref{sec:2.2}, for those safety-critical problems with hard constraints (i.e., constraint needs to be enforced at each time step), the model-based safe layer is more advantageous to ensure safety with the help of model information. However, the cost function in Eq. (\ref{eqn_CMDP_cost}) is a model-free formulation by defining the safety based on the expectation, which is more suitable for complex control tasks with multiple soft constraints in power systems, such as energy/demand management \cite{sayed2023online, zhang2024networked}. 
Essentially, Eqs. (\ref{CMDP_main}) aims to solve a constrained optimization problem, where the main challenge is that both the objective and constraints are non-convex for the RL agent. To effectively transform the original constrained optimization problem into an unconstrained one, commonly used methods in power systems can be generally summarized as follows: the Lagrange multipliers method, the trust region method, the Lyapunov method, and the safety-guided exploration method.

\subsubsection{Lagrange multipliers method}
Lagrangian relaxation is a common solution for constrained optimization problems\cite{bertsekas2014constrained}. In the RL framework, a Lagrange multiplier $\lambda\geq 0$ is introduced to manage a trade-off between the reward and constraint violation costs. Specifically, the original constrained optimization problem in Eqs. (\ref{CMDP_main}) are converted into an unconstrained one as:
\begin{align}
    \min_{\lambda\geq 0}\max_{\pi} \mathcal{L}(\pi,\lambda)
    \dot{=}J_R^\pi - \lambda(J_{C}^\pi - d)
    .\label{eqn_Lagrange_problem}
\end{align}
With the increasing of $\lambda$, the solution of Eq. (\ref{eqn_Lagrange_problem}) converges to the result of the original problem in Eqs. (\ref{CMDP_main}). Note that, during training, the update iteration for the policy $\pi$ is suggested to adopt a faster timescale than that for Lagrange multiplier $\lambda$. That is, as $k$-th iteration, assuming that $\lambda^k$ is constant, the policy $\pi$ is updated for several iterations by solving Eq. (\ref{eqn_Lagrange_problem}) to maximize $\mathcal{L}(\cdot,\lambda^k)$. Then, the $\lambda^k$ is increased in a slower timescale to satisfy the constraint and repeat the iteration process. The update of $\lambda^k$ is set as:
\begin{align}
    \lambda^{k+1} = [\lambda^k - \eta_\lambda(J_{C}^\pi - d)]^+
    .\label{eqn_update_lambda}
\end{align}
where $\eta_\lambda$ is the step size for updating $\lambda^k$, and $[\cdot]^+$ projects $\lambda^k$ into a non-negative real number. Based on the continuous iteration of updating the policy $\pi$ and Lagrange multiplier $\lambda$, this method can guarantee the convergence to a local optimal and feasible solution when the following three assumptions hold\cite{borkar2005actor, bhatnagar2012online}: 
1) $J_{R}^\pi$ is bounded for all policies $\pi$; 
2) every local minimum of $J_{C}^\pi$ is feasible; 
3) $\sum_{k=0}^\infty\eta_\lambda=\sum_{k=0}^\infty\eta_\theta=\infty$ and $\sum_{k=0}^\infty\eta_\lambda^2+\sum_{k=0}^\infty\eta_\theta^2\leq\infty$. 
Here, $\eta_\theta$ denotes the step size of the policy neural network. 
Hence, the design of the step size and initial value for $\lambda$ is significant for the feasibility of the local optimal solution, where the hyperparameter tuning process is also one of the challenges in practice.

The key idea of the Lagrange multipliers method is not complex to implement, so this method has been applied in various control tasks in power grids\cite{8_CSAC_VVC, 25_CSAC_LFC, hu2024multi}. In addition, this method has been similarly extended to different state-of-the-art RL algorithms, such as soft actor-critic (SAC)\cite{SAC_Lag}, trust region policy optimization (TRPO) \cite{TRPO_Lag}, and proximal policy optimization (PPO)\cite{PPO_Lag}. 
However, as shown in Eq. (\ref{eqn_CMDP_constraint}), the constraint safety is defined on all possible states' expectations, leading to a fatal flaw: each specific state is allowed to be unsafe as long as the expectation of states satisfies the constraint. That is, the expectation-based constraint safety cannot prevent some worst cases in safety-critical domains. To address this issue, some researchers recently have tried to introduce a chance constraint\cite{CMDP_PDO} or conditional value-at-risk (CVaR)\cite{WCSAC} to describe the tail risk of constraint violations, for improving the policy robustness. The improved methods can effectively take extreme scenarios into account, according to different risk requirements or preferences.
In practice, the aforementioned methods all require quite strict mathematical assumptions for converging to local saddle points. Then, under mild assumptions, Tessler et al. \cite{CMDP_RCPO} have proposed \textit{reward constrained policy optimization} (RCPO) and proven it can converge almost surely to a constraint-satisfying solution. Besides, the RCPO algorithm is reward agnostic and does not require prior knowledge. The disadvantage of the RCPO is that multiple learning rates are involved, which are difficult to adjust in practice. 
As discussed before, most Lagrange multipliers-based methods can only cope with soft constraints, because they cannot prove to achieve zero constraint violation. To explore whether it is possible to achieve the optimal sublinear convergence rate with zero constraint violations, Bai et al. \cite{bai2022achieving} designed a conservative stochastic primal-dual algorithm (CSPDA) by utilizing the conservative idea of reducing the regret\cite{akhtar2021conservative}, and gave the theoretical convergence analysis.

In summary, the Lagrangian multipliers method transmutes the constrained optimization problem into an unconstrained one by introducing an auxiliary penalty component, thereby enabling the solution to satisfy the constraints and maximize rewards. This method can assure constraint safety as the policy asymptotically converges. Despite its advantages, there are still several limitations: 1) Substantial computation burden for solving a saddle point optimization problem, which equals solving a succession of MDPs; 2) Significant hyperparameter tuning overhead caused by the sensitivity to the initial values and learning rates of the Lagrange multipliers; 3) The convergence rate of the iteration solution cannot be guaranteed, because the objective of the Lagrangian multiplier problem is neither convex nor concave. 

\subsubsection{Trust region method}\label{TR_based}
Different from the penalty of the Lagrange multiplier, the trust region method solves the constrained optimization problem in Eq. (\ref{CMDP_main}) through direct modification of the policy gradient, by enforcing a trust region constraint\cite{TRPO}. Specifically, in the policy iteration, the range of policy parameter changes is limited within a neighborhood of the most recent iterate (i.e., the trust region). This trust region constraint ensures that each step's change is not too large, thereby maintaining the safety and reliability of the policy optimization process. The enforced constraint is formulated as follows:
\begin{subequations}
\label{CPO_main}
    \begin{align}
        \pi_{k+1}=\arg\max_{\pi} 
        &J_R^\pi \\
        \text{s.t.:}~
        &J_{C}^\pi\leq d,\\
        &D(\pi, \pi_k)=||\theta-\theta_k||_2\leq\delta,
    \end{align}
\end{subequations}
where $D$ is the distance measurement; $\delta\geq0$ is a step size; $\theta$ denotes the network parameters of policy $\pi$. At each iteration $k$, solving policy $\pi_{k+1}$ is difficult because it is required to evaluate whether a policy is feasible for trust region constraint. To address this challenge, Achiam et al. \cite{CMDP_CPO} extended the TRPO method and first proposed a general policy search method, called constrained policy optimization (CPO). The key idea of CPO is: firstly, conducting \textit{surrogate} functions to approximate the non-convex objective function $J_R^\pi$; secondly, expanding the objective/cost functions by Taylor second order to simply the problem in Eq. (\ref{CMDP_main}) into a convex optimization problem. 

In CPO, it adopts backtracking techniques to search for new policies, significantly slowing down training efficiency. To improve the efficiency issue, Yang et al.\cite{CMDP_PCPO} introduced two steps (i.e., reward improvement step and projection step) for policy searching, and proposed the projection-based constrained policy optimization (PCPO). To solve the trust region problem, CPO and PCPO both need to calculate the inversion of the Fisher information matrix (FIM). However, when facing high-dimensional policies, calculating FIM becomes impractically expensive, requiring low-cost approximation for FIM. To reduce the approximation error of FIM, Zhang et al. \cite{FOCOPS} proposed first-order constrained optimization in policy space (FOCOPS), whose main idea is to use the primal-dual gradient to solve the trust region problem. Compared with CPO and PCPO, FOCOPS only uses linear approximation and does not need to solve the inversion of FIM, which is more efficient and practical in computation. 

Although the simulation results on the high dimensional continuous control task show that the first-order approximation's performance in FOCOPS is better than that of the complex second-order approximation (e.g., CPO), this observation has not been theoretically substantiated.
Similarly, based on the traditional proximal policy optimization (PPO), Zhang et al.\cite{CMDP_P3O} proposed penalized proximal policy optimization (P3O) algorithm to handle the difficulty of calculating inverse FIM. In P3O, the cost constraint is transformed into an unconstrained optimization problem by the exact penalty function and solved by first-order optimization, which avoids quadratic approximation and high-dimensional Hessian matrix inversion in large CMDP problems.
The aforementioned methods are all trust region methods for solving CMDPs, where the approximated constraints are enforced in every policy update round. Hence, the converged policy can ensure constraint safety during training. Because the above methods are related to TRPO, it is not difficult to apply them to the PPO for constraint optimization. However, it is still not clear how to combine the above methods with RL frameworks that are not of the proximal policy gradient type, such as the RL algorithms with actor-critic framework. 
In addition, there are some limitations to the above methods: 1) The convex approximation of non-convex policy optimization will produce non-negligible approximation errors, so whether the first or second-order approximation can only learn the policy that is close to satisfying the constraints. 2) When the original problem is not feasible under a certain initial policy, we need to restore the policy to the feasible set through interaction with the environment, causing a low sampling efficiency. 3) The second-order approximation involves matrix inversion, which is costly in a high-dimensional environment and is not suitable for solving large-scale CMDP problems.
\begin{figure*}
    \hspace{9em}
    \includegraphics[width=1.7\columnwidth]{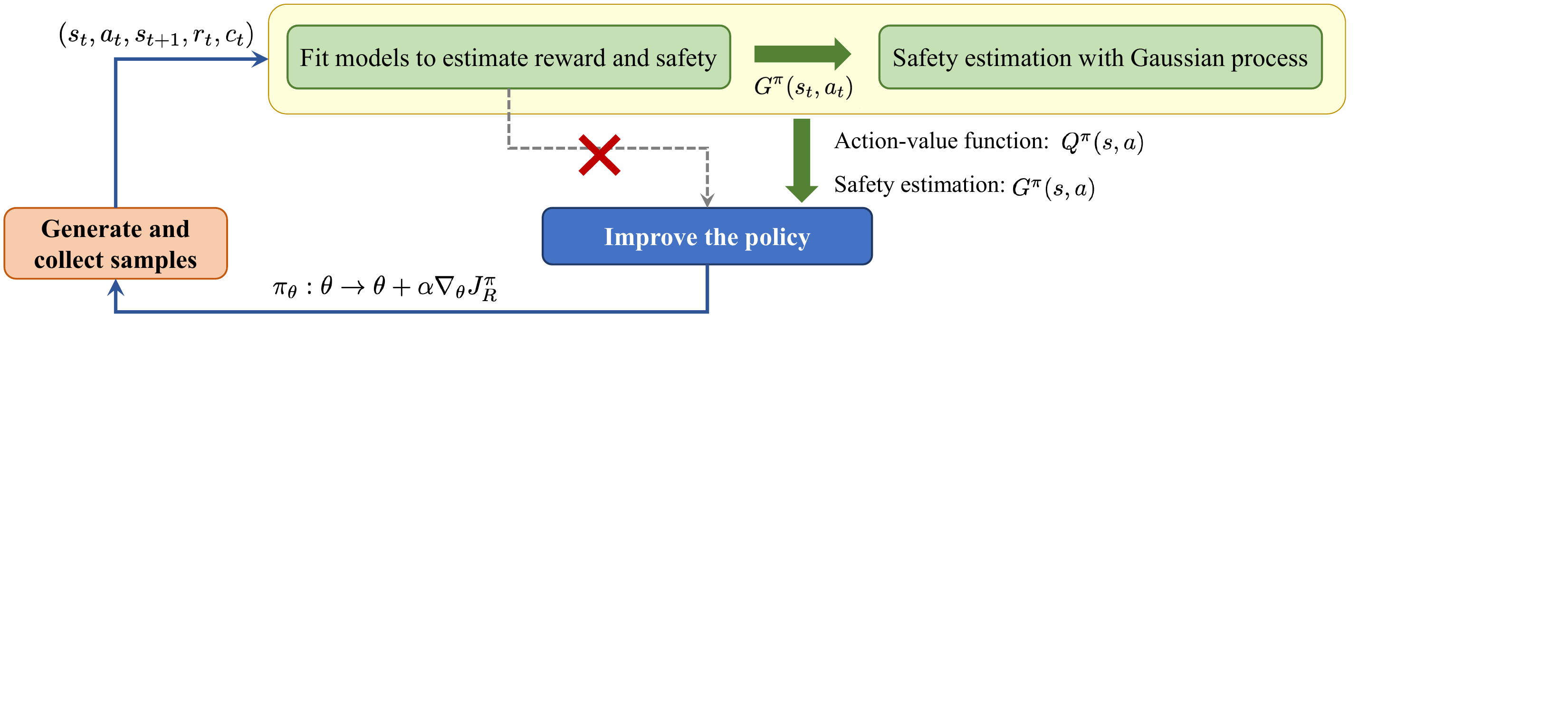}
    \vspace{-39mm}
    \caption{The safety-guided RL framework for safe explorations.}
    \label{fig_safe_guided_scheme}
\end{figure*}

\subsubsection{Lyapunov method}
The Lyapunov function, essentially a concept in the analysis of the system stability, is used to measure the ``distance" of a system state relative to some stable point or set. In the context of safe RL, during the process of exploring, a Lyapunov function can be introduced to ensure the system state is not far from the predefined safe state\cite{perkins2002lyapunov}. Generally, adopting the Lyapunov method for solving a CMDP problem includes three key steps\cite{chow2019lyapunov, jeddi2021lyapunov}: 
First, construct a Lyapunov function $L\in\mathcal{L}_\pi(s_0,d)$ that maps the system state to a real value, which measures the ``distance" between the system state and the predefined safe state. Second, reformulate a constrained optimization problem to add Lyapunov constraints satisfying the defined Lyapunov function. Last, propose a policy update approach to embed the Lyapunov constraints into the policy network. 

However, the most necessary but challenging step in practice is how to design a satisfactory Lyapunov function based on the known system dynamic, where the Lyapunov function is required to possess several properties (e.g., positive definiteness, decay property, and Lipschitz continuity condition)\cite{cui2021decentralized}. Hence, the design of a Lyapunov function requires a deep understanding and analysis of the system dynamic, and can only be combined with model-based RL algorithms. 
In addition, for different systems, it is required to design the system's own specific Lyapunov function that is not universal, especially for complex or highly nonlinear systems. Once an analytic Lyapunov function is designed properly, this method can effectively guarantee the stability of the system with optimal control performance, which is critical and a research hotspot for the RL application in power systems (e.g., frequency control\cite{yuan2024reinforcement, cui2022reinforcement}).

\subsubsection{Safety-guided exploration method}
To cope with the design issue of the Lyapunov function, reference \cite{Safety_guided_Exploration} formulated a model-free function for safety costs as the candidate Lyapunov function, and modeled its derivative with a Gaussian process which provides statistical guarantees. This model-free method steers the policy search in a direction that decreases the safety costs and increases the objective reward, which can effectively solve power system control tasks involving multiple complex systems\cite{qiu2022safe}.

The traditional RL algorithm updates the policy only by estimating the objective reward through the \textit{action-value} function $Q^\pi(s,a)=\mathop{\mathbb{E}}_{\tau\sim\pi} \left[ \sum\nolimits_{t=0}^T\gamma^tr_t(s,a) |_{s_0=s,a_0=a}  \right]$. Hence, the conventional policy gradient direction follows:
\begin{align}
    &\nabla_\theta J_R^\pi = \mathbb{E}_{s\sim\rho^\pi}[\nabla_\theta\pi(s) \nabla_{a}Q(s,a)|_{a=\pi(s)}]
    ,\label{eqn_J_R_gradient}
\end{align}
where $\rho^\pi$ is the state distribution to the policy parameters $\theta$. To achieve safety-guided exploration, a model-free safety estimation $G^\pi(s, a)$ with the Gaussian process is introduced to ensure safety. The safety estimation is defined as $G^\pi(s, a) = \mathbb{E}_{\tau\sim\pi} \left[ \sum\nolimits_{t=0}^T\gamma^t c_t(s, a)|_{s_0=s,a_0=a}  \right]$, which is approximated in practice with a deep neural network. Then, the original policy gradient direction is re-derived considering both the action-value function $Q^\pi(s, a)$ and safety estimation $G^\pi(s, a)$, rewritten as:
\begin{align}
    &\nabla_\theta J_R^\pi = \mathbb{E}_{s\sim\rho^\pi}[\nabla_\theta\pi(s) \nabla_{a}Q(s,a)|_{a=\pi(s)} \nabla_{a}G(s,a)|_{a=\pi(s)}]
    .\label{eqn_J_RC_gradient}
\end{align}
Hence, the whole safety-guided RL framework is that: 1) the collected samples are used to fit models $Q^\pi(s, a)$ and $G^\pi(s, a)$, which estimate the objective reward and safety cost, respectively; 2) the Gaussian process estimation is updated in every iteration; 3) the policy $\pi$ is finally optimized following the rewritten gradient $\nabla_\theta J_R^\pi$ in Eq. (\ref{eqn_J_RC_gradient}) which combines the reward and safety estimations.
However, this method can only combine with the RL algorithm using the actor-critic framework, while not for all RL frameworks. Although the stability certificates of Gaussian process estimation can provide high-probability trajectory-based safety guarantees for unknown environments, how the initial knowledge influences the efficacy of this method remains to be explored.

\section{Applications of safe reinforcement learning in power systems}
\label{sec:3}
\begin{figure*}
    \hspace{1em}
    \includegraphics[width=1.9\columnwidth]{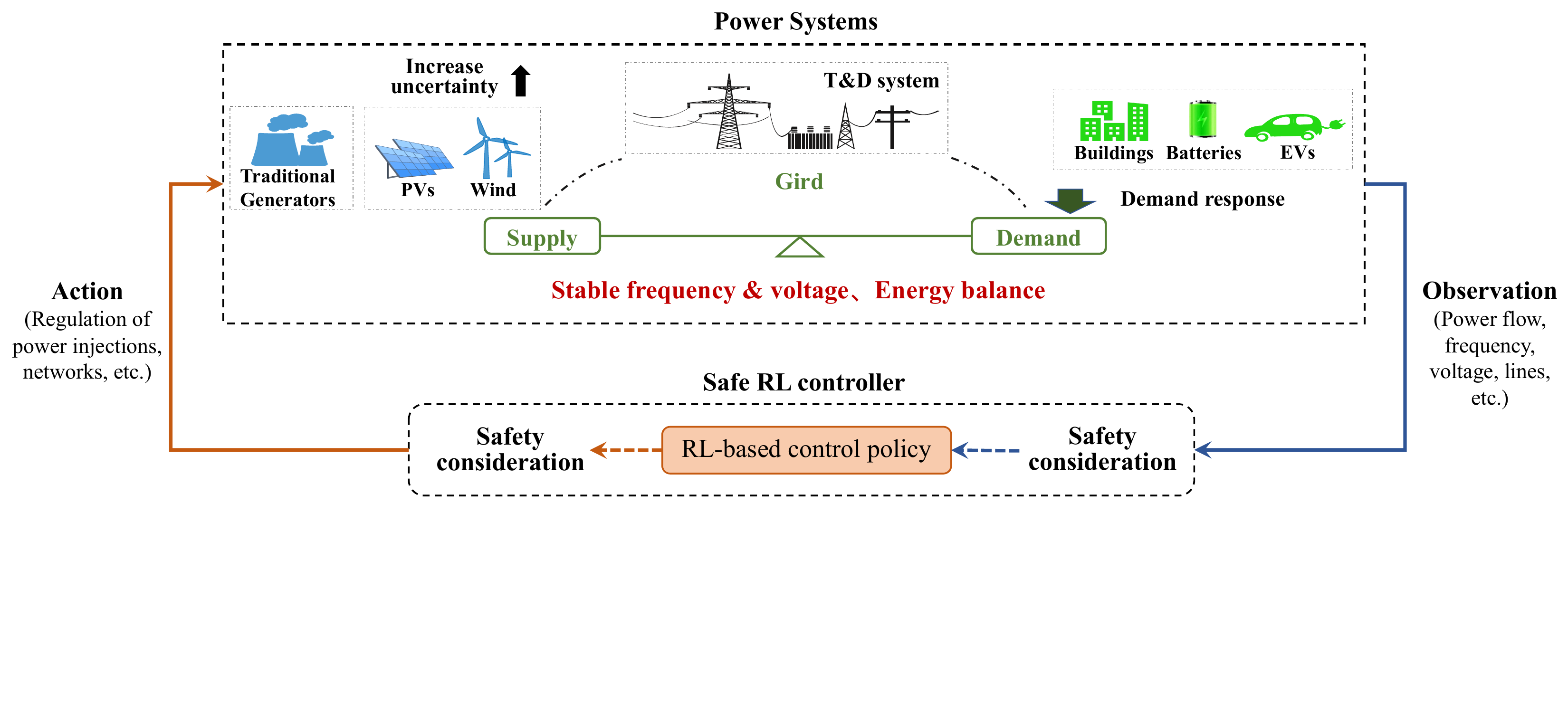}
    \vspace{-19mm}
    \caption{The application of safe RL in power systems.}
    \label{fig_safe_RL_application}
\end{figure*}

With the development of artificial intelligence and Internet of Things technologies, model-free RL-based control methods are widely applied for complex tasks in power systems, to cope with operation environments with high uncertainties. Traditional RL methods rely heavily on large neural networks with millions of parameters, which seem the inexplicable ``black box" and cannot ensure system safety. Hence, for safety-critical problems in power systems, safe RL techniques become an appealing complement to the application of traditional RL. As illustrated in Fig. \ref{fig_safe_RL_application}, the safe RL scheme considers the ``safety" concept before the final decision is executed, which relieves the exploration risk of control decisions but still converges to optimal control policies.
Specifically, in power systems, the constraint safety of the frequency and voltage is the most critical indicator of system operation, and real-time power balance requires reliable energy management\cite{Lina_RL_review}. Therefore, this section focuses on the following three key applications: frequency regulation, voltage control, and energy management, as summarized in Tables \ref{FR_summary}, \ref{VC_summary}, and \ref{EM_summary}, respectively. 
For power systems, frequency regulation is a continuous control problem with hard constraints that must be satisfied, while energy management is usually a sequential decision-making problem with soft constraints. Voltage control probably has both hard and soft constraints according to system scales and types. Facing different scenario requirements, various safe RL techniques are adopted to tackle safety challenges that cannot be solved in traditional RL frameworks. In the following subsections, we elaborate on the detailed design of MDPs and safe modules in different applications.

\subsection{Frequency regulation}
Frequency regulation (FR) is a critical aspect of power system operation and stability. It aims to maintain the system frequency within acceptable limits, essential for preventing blackouts or equipment damage. There are typically three timescales associated with FR control\cite{bevrani2011intelligent}: 
1) Primary control, known as ``droop control," is an automatic response triggered by speed governors, which occurs within seconds after a disturbance; 
2) Secondary control, known as ``automatic generation control" (AGC), operates over a time frame of several minutes after droop control, which is usually achieved by adjusting the setpoints of generators; 
3) Tertiary control, known as ``economic dispatch," operates over a time frame of tens of minutes to hours, which is responsible for optimizing the generation to meet the demand with minimum costs. Hence, the tertiary control does not focus on the stability and safety of the frequency, which can correspond to grid-level power dispatch discussed in Section \ref{Enenrgy_management}.

Many studies have applied RL methods to implement FR control, to cope with the increasing penetration of uncertain renewables\cite{Lina_RL_review}. In practice, the safety of FR control policy is vital for power systems, such as the stability of closed-loop systems and satisfying power flow constraints. However, most traditional RL methods cannot handle the safety issue properly. Therefore, some recent works have improved the RL framework and proposed safe RL techniques for FR control, summarised in Table \ref{FR_summary}. In this subsection, we take secondary control as an example, because it is the most typical FR problem and has been extensively studied \cite{25_CSAC_LFC, xia2022safe, wan2023adapsafe}. 
We will first present the generalized system dynamic model formulation, state/action spaces, and reward/cost functions. Then, the safe RL techniques and key issues in safe RL-based FR are further analyzed. 
\begin{table*}
    \renewcommand\arraystretch{1.4}
    \centering
    \caption{Literature summary of frequency regulation.}
    \setlength{\tabcolsep}{3.8mm}
    \footnotesize
    \begin{tabular}{p{1.2cm}p{2.8cm}p{3cm}p{1.8cm}p{5.5cm}}
        \hline
        \centering Reference&\centering Problem &\centering Constraint&\centering Methodology&Key Features\\
        \hline
        \centering \cite{ cui2022reinforcement} (2021) \newline\cite{cui2021lyapunov} (2022)
        & \centering Primary FR  
        & \centering System stability
        & \centering Safe layer, Lyapunov function
        & Train the neural Lyapunov function to satisfy the positive definiteness of its value and the negative definiteness of its Lie derivative.\\
        \hline
        \centering\cite{25_CSAC_LFC} (2022)
        & \centering Load frequency control    
        & \centering Line power flows
        & \centering CSAC, Lagrange method
        & Restrict the entire policy space to a smaller allowable safe space using the Lagrange multiplier method.\\
        \hline
        \centering\cite{xia2022safe} (2022)
        & \centering Multi-area microgrid FR
        & \centering Operation constraints for generators and ESSs
        & \centering NN-based safe layer
        & Propose a safe module consisting of two components: a safety evaluation network and an action guidance network.\\
        \hline
        \centering\cite{lesage2022batch} (2022)
        & \centering Thermostatically controlled loads providing FR
        & \centering voltage magnitude limits
        & \centering Offline(batch) RL
        & Leverage historical network measurements to train the offline RL controller, and reduce voltage magnitude constraint violations.\\
        \hline
        \centering\cite{zhao2023barrier} (2023)
        & \centering General solution to control problems, e.g., FR
        & \centering Frequency threshold 
        & \centering Shielding, CBF
        & Complement the RL action by the optimal solution guaranteeing the safety of systems, based on a Gaussian process model.\\
        \hline
        \centering\cite{yu2023district_appl} (2023)
        & \centering Thermostatically controlled loads providing FR  
        & \centering Regulation performance score and users' temperature comfort
        & \centering Safe layer, CBF  
        & 1. Utilize previous CBF controllers to avoid repeatedly taking unsafe actions; 2. Propose a neural network-based method to achieve high-efficiency calculation.\\
        \hline
        \centering\cite{wan2023adapsafe} (2023)
        & \centering Load frequency control  
        & \centering Frequency stability and regulation bound of renewable energy sources 
        & \centering Safe layer, CBF
        & Design a self-tuning CBF-based compensator to realize the optimal safety compensation under different risk conditions.\\
        \hline
        \centering\cite{yuan2024reinforcement} (2024)
        & \centering Transient frequency control
        & \centering Stability condition
        & \centering Safe layer, Lyapunov function
        & Define the search space of distributed control policies that guarantee asymptotically stable and transient-safe closed-loop systems. \\
        \hline
    \end{tabular}
    \label{FR_summary}
\end{table*}

\subsubsection{State and action spaces}
When a generator outage occurs, the system frequency dynamics reflect the relationship between the power balance and the frequency fluctuation $\Delta f$, which can be expressed as:
\begin{align}
    & 2H\frac{d}{dt}\Delta f + D\Delta f = \Delta P^\text{gap},
\end{align}
where $H$ and $D$ are the system inertia constant and load damping constant, respectively;  $\Delta P^\text{gap}$ denotes the imbalanced power gap in power systems. It can be seen that the power imbalance is the key factor influencing the system frequency, so the \textit{action} is defined as the power regulation of the traditional generation and renewable energy resources (RENs) by:
\begin{align}
    &a^\text{F} = \left[  P^\text{gen}_k, P^\text{ren}_i | k\in\mathcal{N}_\text{G},i\in\mathcal{N}_\text{REN} \right]^\intercal,
\end{align}
where $\mathcal{N}_\text{G}$ and $\mathcal{N}_\text{REN}$ are the number of controllable generators and RENs, respectively; $P^\text{gen}_k$ and $P^\text{ren}_i$ are command output power of each generator and REN. Generally, RESs are connected to the power system through power electronic devices, thus the action space is continuous whose scale increases with more generators/RENs being controlled. The design of the \textit{state} captures the current frequency change and power outputs as: $s^\text{F} = \left[  \Delta f, P^\text{gen}_k, P^\text{ren}_i | k\in\mathcal{N}_\text{G},i\in\mathcal{N}_\text{REN} \right]^\intercal$. If other power injections can be controlled for regulation services, such as demand-side loads and energy storage systems (ESSs), their output powers can also be included in the action space\cite{xia2022safe}. 

\subsubsection{Reward and cost functions}
The reward function design determines the control objective of the training RL agent, which is significant for successful RL applications. For FR control problems, the key aim is to maintain the frequency at a nominal value after disturbances. Hence, to maximize the expected reward, the minus of frequency deviation is commonly used to design the reward as:
\begin{align}
    r^\text{F} &= - \Delta t \cdot \sum\nolimits_{i\in\mathcal{N}_\text{B}} |f_i - f^\text{ref}|,
    \label{r_FR}
\end{align}
where $\mathcal{N}_\text{B}$ is the number of the bus; $f_i$ and $f^\text{ref}$ are the measured frequency at bus $i$ and the nominal value, respectively. For multi-area FR, the tie-line power flow is also required to be considered in Eq. (\ref{r_FR}), introducing weight factors for two or more items\cite{yan2020multi}. In addition, the generation regulation cost\cite{li2020deep}, the large penalty for crossing deviation limits\cite{wan2023adapsafe}, and the square or exponential form for deviation frequency\cite{rozada2020load} are also an alternative in reward functions. There is no definitive assertion about which reward function is better, and the reward design is contingent upon the specific scenario.

The constraints for FR mainly come from two aspects, one is the safe range for fluctuated frequency and the other is the acceptable regulation range of control objects, which can be generalized as:
\begin{subequations}
\label{FR_constraint}
    \begin{align}
        &\underline{f} \leq f\leq \overline{f}, \\
        &\underline{P}^\text{gen}\leq  P^\text{gen}_k \leq \overline{P}^\text{gen},\quad\forall k\in\mathcal{N}_\text{G},\\
        &\underline{P}^\text{REN}\leq  P^\text{REN}_i \leq \overline{P}^\text{REN},\quad\forall i\in\mathcal{N}_\text{REN},
    \end{align}
\end{subequations}
where $\overline{f}$ and $\underline{f}$ denote the upper and lower bounds of system frequency; $\overline{P}^\text{gen}/\overline{P}^\text{REN}$ and $\underline{P}^\text{gen}/\underline{P}^\text{REN}$ represent the upper and lower bounds of output power in generator and RENs, respectively. If more types of controllable objects are considered, there shall be more physical operating constraint limits, e.g., state-of-charge and charging rate in ESSs\cite{xia2022safe}, comfort requirement in demand-side loads\cite{yu2023district_appl}, etc. Then, the cost function is the total penalty for all the constraint violations:
\begin{align}
    & c^\text{F} = \sum\nolimits_{m} \beta^\text{m}\cdot
    [\max(0, \underline{b}^\text{m}-b^\text{m}) + \max(0, b^\text{m}-\overline{b}^\text{m})]^2,
\end{align}
where $m$ is the constraint number in Eq. (\ref{FR_constraint}); $\beta_\text{l}$ is the corresponding weight factor; $b^\text{m}$, $\underline{b}^\text{m}$, and $\overline{b}^\text{m}$ are formal variables to denote the specific three variables, three lower limits, and three upper limits in Eq. (\ref{FR_constraint}). If primary and secondary control processes are both considered, the piecewise function  is also an alternative to formulate the cost function\cite{wan2023adapsafe}.

\subsubsection{Safe RL techniques}
Apart from batch RL (i.e., offline RL), most published studies choose to apply safe layer-based approaches in FR problems, because a safe layer needs to check the system safety at every time step, which is more effective for hard and critical constraints. The specific design of the safe layer in FR mainly includes three ideas:
(1) With an accurate constraint model, the next state is directly and mathematically derived to judge whether the state is safe. Reference \cite{cui2022reinforcement} used the physical system model to derive the structure of the stabilizing RL controller based on Lyapunov theory.
(2) With a poor constraint model, the system model is approximated through the Gaussian process or other methods, then the safe area is defined by proposed barrier functions. For instance, reference \cite{wan2023adapsafe} proposed an adaptive and safe-certified RL algorithm for FR control, where the safe layer is designed based on Gaussian process regression and CBF-based compensator. Reference \cite{zhao2023barrier} proposed a more general solution for power system control problems (including FR), which certifies the neural barrier function that perseveres barrier conditions. This method does not rely on the model approximation as the first step.
(3) Without any model information, an NN-based safety layer is introduced to judge the state safety and correct the action. More specifically, reference \cite{xia2022safe} proposed a safety evaluation network as a safety monitor, and an action guidance network for safe action-guiding. In this manner, the training of the proposed safe layer relies on the dataset instead of the physical model.

\subsubsection{Discussion}\leavevmode
 
\textbullet \textit{ Stability:}
The integration of RL agents with traditional existing controllers requires the closed-loop system dynamics to be stable, which is different from constraint safety and rarely considered in the current works. Cui et al. \cite{cui2021lyapunov} have tried to propose a Lyapunov regularized RL approach for FR control to cope with the transient stability, where the Lyapunov function is parameterized through NNs. However, this work only focuses on stability ignoring constraint safety, so combining stability and safety is a potential research direction for RL applications in FR control.

\textbullet \textit{ Demand-side regulation services:}
The aforementioned design of state and action spaces mainly focuses on controlling generators to regulate the power supply. With increasing demand-side resources participating in electrical markets, ubiquitous load devices are also an alternative as a compensatory to FR control. Reference \cite{yu2023district_appl} proposed to control large-scale district cooling systems for FR services, which satisfy users' thermal comforts by introducing CBF in safe layers. Besides, in \cite{lesage2022batch},  thermostatically controlled loads are controlled through a batch RL method to provide FR, which ensures constraint safety by offline training. More potential load control for FR services remains to be explored.

\subsection{Voltage control}
With the increasing integration of renewable energy sources, the highly uncertain renewables have brought rapid voltage fluctuations into distribution grids. The voltage control problem in power systems aims to keep network voltages within an allowable range and reduce system losses, by determining the control actions for all voltage regulating and var control devices.  Many published works have developed model-free RL methods for voltage control to tackle the challenges of renewables' uncertainties and unknown accurate system models\cite{gao2021consensus, wang2020data}. However, traditional RL applications do not have an explicit mechanism to ensure the safety of operation constraints, e.g., voltage limits and line flow limits. This is because most RL controllers are defined by large-scale neural networks (NNs), which is considered as a ``black box" and trained through the feedback of ``trial and error". 
To this end, considering constraint safety, some recent studies have tried to explore a safe model-free RL framework for voltage control, which is summarised in Table \ref{VC_summary}. We will present the most commonly used definitions of state/action spaces, reward/cost functions, and the application of safety techniques.
\begin{table*}[t]
    \renewcommand\arraystretch{1.4}
    \centering
    \caption{Literature summary of voltage control.}
    \setlength{\tabcolsep}{3.8mm}
    \footnotesize
    \begin{tabular}{p{1.2cm}p{2.8cm}p{3cm}p{1.8cm}p{5.5cm}}
        \hline
        \centering Reference& \centering Action Space &\centering Constraint&\centering Methodology&Key Features\\
        \hline
        \centering \cite{8_CSAC_VVC} (2020)
        & \centering Set tap on/off position
        & \centering Nodal voltage limits
        & \centering CSAC, Lagrange multiplier method
        & Use an ordinal network structure to encode the natural ordering between actions of voltage regulating devices, and the inductive bias is introduced to accelerate the learning.\\
        \hline
        \centering \cite{kou2020safe} (2020)
        & \centering Reactive power adjustments of hybrid distribution transformer 
        & \centering Voltage state constraint
        & \centering Model-based safe layer
        & Use sensitivity matrix to predict the bus voltages change, and compose an additional safety layer by a convex problem on top of the RL framework.\\
        \hline
        \centering \cite{vu2021barrier} (2021)
        & \centering Load shedding ratio 
        & \centering Voltage magnitude limits
        & \centering Reward shaping with CBF
        & A well-designed barrier function is included in the reward function to guide the learning without model knowledge.\\
        \hline
        \centering \cite{cui2021decentralized} (2021)
        & \centering Nodal reactive power
        & \centering Voltage stability (Lipschitz constraint)
        & \centering Engineer the NN structure of RL controllers
        & Optimize the set of Lipschitz bounds to enlarge the search space of RL controllers, and  train locally at each bus with policy gradient algorithm.\\
        \hline
        \centering \cite{vu2021safe} (2021)
        & \centering Load shedding amount
        & \centering Voltage recover criteria
        & \centering Lagrange multiplier method
        & Formulate a Lagrangian function involving both the normal reward function and the safety function with a multiplier.\\
        \hline
        \centering \cite{ li2022learning} (2022)
        & \centering Battery storage systems, capacitor banks, OLTCs, voltage regulators 
        & \centering Substation/BSS capacity constraints and nodal voltage/branch loading limits
        & \centering CPO
        & 1. Design a stochastic policy to handle mixed discrete and continuous action space; 2. Employ the CPO algorithm to handle the operational constraints.\\
        \hline
        \centering \cite{gao2022model} (2022)
        & \centering OLTCs, voltage regulators 
        & \centering Voltage magnitude limits
        & \centering NN-based safe layer
        & Propose a quadratic programming-based safe layer based on neural network architecture to enhance the safety.\\
        \hline
        \centering \cite{hossain2023efficient} (2023)
        & \centering Load shedding of each bus
        & \centering Voltage recovery criteria
        & \centering Offline(batch) RL 
        & Use a surrogate model to generate roll-outs in the RL training stage, and add imitation learning to reduce the early unsafe exploration of the policy training. \\
        \hline
        \centering \cite{jeon2023safety} (2023)
        & \centering Reactive power of mobile ESSs and PV smart inverters
        & \centering State of charge limits, voltage magnitude limits
        & \centering Model-based action replacement
        & Propose two safety modules with plug-and-play functionality to ensure a safe exploration, based on the system physical model. \\
        \hline
        \centering \cite{guo2023safe} (2023)
        & \centering Reactive power injections of PV inverters
        & \centering Nodal voltage limits and power flow limits
        & \centering Model-based safe layer
        & A safety projection is added to analytically solve an action correction formulation per each state, which guides the exploratory actions in the direction of feasible policies.\\
        \hline
        \centering \cite{yan2023multi} (2023)
        & \centering Voltage magnitude, local loads, active and reactive power outputs from PV inverters
        & \centering Voltage magnitude limits
        & \centering Lagrange method, graph convolutional networks (GCN)
        & 1. Use the GCN module to denoise graph signals and improve robustness against flawed input data; 2. Transform the original problem into an unconstrained min-max problem by Lagrange multipliers. \\
        \hline
        \centering \cite{sun2023optimal} (2024)
        & \centering Reactive power injections of PV inverters
        & \centering Voltage magnitude limits and voltage unbalance 
        & \centering Human intervention
        & An online switch mechanism between action exploration and human intervention is designed to facilitate the learning process towards human actions.\\
        \hline
    \end{tabular}
    \label{VC_summary}
\end{table*}

\subsubsection{State and action spaces}
For voltage control in distribution systems, different controllable devices possess different operational characteristics, which can be classified into discrete control and continuous control. Discrete control devices include switchable capacitor banks (SCBs), on-load tap changing transformers (OLTCs), and voltage regulators (VRs), 
which are controlled in a slow timescale on an hourly or daily basis. Continuous control devices, such as battery storage systems (ESSs), distributed resources (DRs), and static Var compensators (SVCs), can regulate their active/reactive power in a fast timescale by seconds.
Since the voltage control problems involve both the distribution network and controllable devices, the observed system state $s$ shall capture the significant information of both above two sectors, which can be defined as:
\begin{align}
    s^\text{Vol}=
    &[ t, (P_i,Q_i,v_i)_{i\in\mathcal{N}_\text{B}}, 
    \tau^\text{SCB}, \tau^\text{VR}, \tau^\text{OLTC},\notag\\
    &P^\text{DG}, Q^\text{DR}, Q^\text{SVC}, E^\text{BSS}]^\intercal,
\end{align}
where $\mathcal{N}_\text{B}$ is the number of the bus. State $s_t$ consists of four parts: (1) the time information of current time step $t$; (2) the network information of net active power $P_i$, net reactive power $Q_i$, and voltage magnitude $v_i$ at bus $i$; (3) the device information of the discrete tap positions of SCBs/OLTCs/VRs by $\tau^\text{SCB}$, $\tau^\text{VR}$, and $\tau^\text{OLTC}$, respectively; (4) the device information of the active power outputs of DGs $P^\text{DG}$, reactive power outputs of DGs/SVCs by $Q^\text{DG}$, $Q^\text{SVC}$, and stored energy in ESSs by $E^\text{BSS}$.

The action space of the voltage control problem depends on the selected controllable devices. When all the aforementioned devices are considered, the RL agent's action $a$ can be defined as:
\begin{align}
    &a^\text{Vol}=\left[ \Delta \tau^\text{SCB}, \Delta \tau^\text{VR}, \Delta \tau^\text{OLTC},
    P^\text{DG}, Q^\text{DG}, Q^\text{SVC}, P^\text{BSS} \right]^\intercal,
\end{align}
where $\Delta \tau^\text{SCB}$, $\Delta \tau^\text{VR}$, $\Delta \tau^\text{OLTC}$ denote the discrete changes of corresponding tap positions. Parameter $P^\text{BSS}$ represents the control variable of a BSS, where a positive value means the BSS is charging and a negative value represents discharging. The control variables of continuous control devices (i.e., DGs and SVCs) are exactly their active/reactive power outputs.
Since the mixed discrete and continuous action space cannot be handled by traditional RL methods directly, most published safe RL techniques in voltage control only consider either continuous devices (e.g., inverter-based DGs) \cite{guo2023safe, yan2023multi} or discrete devices (e.g., OLTCs, VRs) \cite{8_CSAC_VVC, gao2022model}. Nevertheless, the recent studies \cite{li2022learning,liu2021bi} propose safe RL for a hybrid action space, taking into account both the safety of discrete and continuous actions simultaneously.

\subsubsection{Reward and cost functions}
The reward design directly guides the achievement of the voltage control objective. Since a key objective of voltage control is to reduce network losses, the negative of the total network loss in the distribution network is the most commonly used reward\cite{guo2023safe, li2022learning, yan2023multi}. On this basis, considering the device loss brought by frequent switching (especially for slow timescale devices), some researchers also introduce an extra penalty item to reduce control efforts\cite{kou2020safe, gao2022model}.
Hence, the commonly used reward function can be generalized as:
\begin{align}
    &r^\text{Vol} = - \beta C^\text{loss}\sum\nolimits_{i\in\mathcal{N}_\text{B}} P_i - (1-\beta)C^\text{reg} \sum\nolimits_{|\mathcal{A}|}||\Delta a||_2,
\end{align}
where $\beta$ is the weight factor; $C^\text{loss}$ and $C^\text{reg}$ are cost coefficients of network loss and device regulation; $\Delta a$ denotes the change of device state after executing actions. Note that, in reward functions, reducing the tap position changes and power regulations is only considered as part of the objective rather than a hard constraint. The safety-critical constraints are usually designed as the following cost function to ensure constraint safety.

The key constraint for voltage control is to limit the voltage magnitudes within an acceptable range or stay at a nominal value. Based on the required upper and lower limits $\overline{v}$, $\underline{v}$ of nodal voltages, the cost function is typically defined as the voltage violation value. For example, in \cite{yan2023multi, gao2022model, kou2020safe, 8_CSAC_VVC}, the cost function $c$ is formulated in a generalized way as:
\begin{align}
    & c^\text{Vol} = \sum\nolimits_{i\in\mathcal{N}_\text{B}}(|v_i-\overline{v}|+|v_i-\underline{v}|-|\overline{v}-\underline{v}|).
\end{align}
Moreover, some recent studies have begun to take more system operation constraints into account, such as the branch power flow and thermal current constraints \cite{guo2023safe}, substation/BSS capacity constraints\cite{li2022learning}, and other necessary sophisticated constraints\cite{jeon2023safety}.

\subsubsection{Safe RL techniques}
Both the safe layer-based and policy optimization-based methods have been applied for voltage control:

\textbullet \textit{ Safe layer:}
As described in Table \ref{VC_summary}, for the safe layer-based method, some works design model-based quadratic programming as the safe layer\cite{jeon2023safety, guo2023safe, kou2020safe}. For instance, in \cite{guo2023safe}, constraints of the safe layer require accurate knowledge of the branch power flow model, to solve safe action projection. In \cite{kou2020safe}, the next voltage state is represented by a first-order linear approximation to obtain an explicit sensitivity matrix, which can analytically describe how much the bus voltages change after executing actions. Moreover, an alternative is to utilize NNs to approximate the unknown system model in the safe layer\cite{cui2021decentralized, gao2022model}, when there is no straightforward analytical solution to the quadratic problem with multiple inequality constraints. 
Two key challenges of the safe layer-based method in voltage control include 1) the next voltage state is hard to predict without models, which hinders the judgment of constraint safety; 2) a well-designed safe layer is difficult to generalize to another distribution network, especially when state/action spaces changes.

\textbullet \textit{ Policy optimization:}
Compared with the safe layer, the CMDP-based policy optimization is model-free and more universal to various distribution networks, which only requires the collection of historical cost value. Typically, the cost function $J_C^\pi$ can be formulated as a penalty term in the objective via a Lagrange multiplier\cite{8_CSAC_VVC, vu2021safe, yan2023multi}, whose limitation is that manually tuning the Lagrange multiplier $\lambda$ usually requires a tedious process of trial-and-error. 
To solve the CMDP in a completely self-adaptive way, the average KL-Divergence $D_\text{KL}(\pi_{t-1},\pi_t)$ is introduced in TRPO-class algorithms to measure the searching area of policy, which releases the requirement on system models\cite{li2022learning}. Then, an advantage function $A^\pi(s, a)=r(s, a, s')+\gamma V(s')-V(s)$ is introduced to formulate the constraint bounds of cost $J_c^\pi$, leading to a convex quadratic optimization that can be solved analytically with a guarantee of global optimum. Nonetheless, the limitation is the huge computation burden to obtain the analytical solution of the optimization, such as calculating the inversion of the FIM.

\subsubsection{Discussion}
Some key issues of applying safe RL to voltage control are discussed below.

\textbullet \textit{ Distributed Multi-device Collaboration:}
As the network scale increases, centralized voltage control requires a central controller leading to a heavy communication burden, which is vulnerable to single-point failures. Moreover, the controllable devices for voltage control may belong to different entities, involving data privacy and multi-device collaboration efficiency. Hence, the commonly used ``centralized single-agent safe RL" is challenged by the urgent concerns of communication failure, privacy, and scalability. Reference \cite{yan2023multi} proposed a multi-agent safe RL to reduce the necessity for communication, based on decentralized partially observable MDPs. Reference \cite{9761229} used federated learning to cope with data privacy in voltage control. However, handling both the scalability and privacy in safe RL should be further investigated, especially when a multi-agent collaboration is required for large-scale networks.
  
\textbullet \textit{ Initial Data Collection:}
Voltage control involves the complex network and power flow, thus current safe RL techniques mainly design safe layers and policy optimization criteria through dynamic neural network approximation. This could still be problematic in real-world power systems because NN-based approximation is unreliable at the early stage, which takes effect when collecting enough data. Two possible remedies are 1) training the policy on digital twin simulators to collect initial data, and 2) applying transfer learning or sim-to-real techniques\cite{bossens2023robust} to generate initial sample data.

\textbullet \textit{ System Operation Constraint:}
Apart from the voltage magnitude limits, more system operation constraints should be taken into account in practice, such as power factor/branch line constraints and device capacity. Although reference \cite{guo2023safe} formulated power flow constraints into the optimization problem in the safe layer, it requires an accurate network operation model, which is impractical. Reference \cite{li2022learning} has tried to consider all constraints in one cost function, while it is difficult to balance the trade-off between different constraints. One may use hierarchical RL to assign different constraints to different layers or sub-tasks to extend the safe RL satisfying multiple operation constraints.

\subsection{Energy management}\label{Enenrgy_management}
Energy management is a broad research area in power systems that aims to maintain the power balance in an economical and reliable manner, which is significant for the large-scale integration of distributed energy resources and a decarbonized future. To this end, the concept of energy management systems (EMSs) is proposed to achieve real-time system control and optimization. However, in practice, stochastic user demands and weather-relied renewables supply cause significant uncertainties. Thus, among massive optimization approaches, model-free RL is paid more attention nowadays in EMSs because it can handle highly uncertain environments without prior model knowledge. Unfortunately, energy management problems usually involve multiple control devices with several local operation constraints and global power balance requirements. It is difficult for conventional RL to capture the physical constraints, destroying secure system operation. Therefore, as summarised in Table \ref{EM_summary}, many recent works explore safe RL techniques for energy management. 
In the rest of this subsection, we classify the research about EMSs into three categories according to the control objects, and introduce the corresponding state, action, constraints, and reward function, respectively. Then the applied safe RL techniques are reviewed based on different frameworks.
\begin{table*}[t]
    \renewcommand\arraystretch{1.4}
    \centering
    \caption{Literature summary of energy management.}
    \setlength{\tabcolsep}{3.8mm}
    \footnotesize
    \begin{tabular}{p{1.2cm}p{2.8cm}p{3cm}p{1.8cm}p{5.5cm}}
        \hline
        Reference& \centering Problem &\centering Constraint&\centering Methodology&Key Features\\
        \hline
        \centering \cite{li2019constrained} (2019)
        & \centering EV charging scheduling
        & \centering Energy requirement
        & \centering CPO
        & 1. Use a DNN to learn the constrained optimal policy in an end-to-end manner; 2. Employ the CPO algorithm to ensure safety.\\
        \hline
        \centering \cite{zhang2019building} (2019)
        & \centering Building HVAC scheduling
        & \centering Temperature comfort
        & \centering Safe layer, MPC
        & Limit the actions within a safe range and the maximum absolute change of actions according to prior knowledge.\\
        \hline
        \centering \cite{zhang2020multi} (2020)
        & \centering Microgrid power management
        & \centering Voltage/current flow limits and other operational limits
        & \centering Model-based policy gradient
        & Employ the gradient information of operational constraints to generate safe and feasible decisions.\\
        \hline
        \centering \cite{li2021online} (2021)
        & \centering Microgrid energy management
        & \centering Voltage/line/ESS/power flow constraints
        & \centering CPO
        & Employ CPO algorithm to train an NN-based policy to achieve constraint safety.\\
        \hline
        \centering \cite{huang2022mixed} (2022)
        & \centering Residential energy management
        & \centering Temperature comfort
        & \centering NN-based safe layer
        & Propose a prediction model-guided safe layer through an online prediction model to evaluate output actions.\\
        \hline
        \centering \cite{qiu2022safe} (2022)
        & \centering Energy hub
        & \centering Electricity power limit and the heat balance constraint
        & \centering Safety-guided exploration
        & Add a safety-guided network to avoid physical constraint violations without adding a penalty term to the reward.\\    
        \hline
        \centering \cite{yan2022hybrid} (2022)
        & \centering Optimal power flow
        & \centering Generator/bus voltage/line flow limits
        & \centering Primal-dual method
        & Combine the primal-dual RL algorithm and power system models to approximate actor gradients by the Lagrangian.\\
        \hline
        \centering \cite{du2022deep} (2022)
        & \centering Microgrid energy dispatch
        & \centering Power flow constraint and generator power limits
        & \centering Offline RL
        & Propose a two-stage learning framework: 1) a pre-training stage with imitation learning; 2) an online training stage with action clipping and expert demonstrations.\\
        \hline
        \centering \cite{ceusters2023adaptive} (2023)
        & \centering Multi-energy management system
        & \centering Electrical and thermal power output constraints
        & \centering Safe layer by action replacement 
        & (1) Propose the safe layer and safe fallback policy to increase the policy's initial utility; (2) Introduce self-improving hard constraints to increase the accuracy of constraints.\\
        \hline    
        \centering\cite{wang2023secure} (2023)
        & \centering Multi-energy microgrids 
        & \centering Constraints and limits of power and gas networks
        & \centering Physical-informed safety layer
        & Learn a security assessment rule to form a safety layer and mathematically solve an action correction formulation. \\ 
        \hline
        \centering\cite{yu2023district} (2023)
        & \centering District cooling system power dispatch
        & \centering Service performance for power reduction and temperature comfort
        & \centering Model-based safe layer
        & Design a partial model-based safe layer based on a linear program to achieve safety-imposing projection.\\
        \hline
        \centering \cite{yi2023real} (2023)
        & \centering Optimal power flow
        & \centering Generation/line/voltage limits
        & \centering Knowledge-data-driven safety layer
        & Propose a model-based safety layer with prior knowledge and updated continuously according to the latest experiences.\\
        \hline
        \centering \cite{ye2023safe} (2023)
        & \centering Microgrid energy management
        & \centering Power balance, line capacities, nodal voltage magnitudes
        & \centering IPO, CBF
        & Employ the IPO algorithm utilizing a logarithmic barrier function to govern the satisfaction of the safety constraints.\\
        \hline
        \centering \cite{sayed2023online} (2023)
        & \centering Integrated electric-gas system
        & \centering Constraints for electricity and gas networks
        & \centering CSAC, Lagrange method
        & Add a safety network to update the constraint violation penalties to guide the policy in a safe direction.\\
        \hline
        \centering \cite{zhang2024networked} (2024)
        & \centering Demand management in distribution network
        & \centering Carbon emission allowances
        & \centering Multi-agent CPO
        & Proposes a consensus multi-agent CPO approach to satisfy the carbon emission limit and preserving private information.\\
        \hline
        \centering \cite{hu2024multi} (2024)
        & \centering Community integrated energy system
        & \centering Constraints for retail energy prices/integrated energy balance
        & \centering Primal-dual, Lagrange method
        & Employ a Lagrangian multiplier to penalize violation cost, using DNN to estimate the policy and action-value function.\\
        \hline
    \end{tabular}
    \label{EM_summary}
\end{table*}

\subsubsection{State and action spaces}
Considering different control objects have different device characteristics, we summarize the design of state, action, and constraints from the following two categories: integrated electricity-gas energy system, and grid-level power dispatch.

\textbullet \textit{ Residential energy management:}
Following the price-based demand response programs (e.g., time-of-use), users are motivated to make optimal schedules of domestic appliances, to minimize electricity costs through residential EMSs. Generally, for end-users, some loads that are essential and cannot be scheduled are considered ``non-shiftable loads", e.g., lighting, television, microwave, refrigerator, etc. Some loads that can be scheduled at different periods but cannot be interrupted are considered ``shifted and non-interruptible loads", such as washing machine and dishwasher. Other flexible loads whose power can be regulated continuously are considered ``controllable loads", where the most common applications are heating, ventilation, and air conditioning (HVAC) and electric vehicles (EVs).
The action space is defined for shifted and non-interruptible loads $i\in\mathcal{N}_\text{NI}$ and controllable loads $j\in\mathcal{N}_\text{C}$, as follows:
\begin{align}
    &a^\text{Res} = [(x_i^\text{NI}, P_j^\text{C})|i\in\mathcal{N}_\text{NI}, j\in\mathcal{N}_\text{C}]^\intercal,
\end{align}
where $x_i^\text{NI}$ is a binary decision variable for shifted and non-interruptible load $i$ (i.e., 1/0 denotes ``on/off”), and $P_j^\text{C}$ is the power of controllable load $j$. Thus, the action space includes both discrete and continuous control variables. 
The state space shall reflect the operation status of domestic devices in residential EMSs, which is usually defined by a high dimensional vector:
\begin{align}
    &s^\text{Res} = [t, \textbf{e}^\text{out}, x_i^\text{NI}, \textbf{o}_i^\text{NI}, P_j^\text{C}, \textbf{o}_i^\text{C})|i\in\mathcal{N}_\text{NI}, j\in\mathcal{N}_\text{C} ]^\intercal,
\end{align}
which is composed of the current time $t$, operating state/power of different devices $x_i^\text{NI}$ and $P_j^\text{C}$, and other related states of environments/shifted and non-interruptible loads/controllable loads, i.e., $\textbf{e}^\text{out}$, $\textbf{o}_i^\text{NI}$, and $\textbf{o}_i^\text{C}$. For instance, the outdoor temperature, users' comfort preference, and electricity price can be contained in $\textbf{e}^\text{out}$ \cite{li2020real}; $\textbf{o}_i^\text{NI}$ can include the remaining time required for devices to complete the task and the task deadline \cite{huang2022mixed}; and $\textbf{o}_i^\text{C}$ captures temperature deviations for HVAC loads or depart time for EVs\cite{li2019constrained}.

Constraints in the residential energy management problem are mainly from human comfort, such as whether all task processes for non-interruptible loads can be finished, the thermal comfort in HVACs, and the charging target of EVs at the departure time. If we use $C^\text{Res}$ to denote the generalized constraints in residential EMSs, there are two types of constraints:
\begin{subequations}
    \begin{align}
        &C^\text{Res}_t = C^\text{Target} ,\quad  \text{If } t=T, \\
        &\overline{C}^\text{Res}\leq C^\text{Res}_t \leq \underline{C}^\text{Res}, \quad \forall 0 \leq t \leq T,
    \end{align}
\end{subequations}
where the first constraint type is to check the completion of tasks through the target $C^\text{Target}$ at the end of the day; and the second constraint type is to limit the upper/lower bounds by $\overline{C}^\text{Res}$ and $\underline{C}^\text{Res}$ during the whole management process. Because each managed domestic appliance has its corresponding comfort requirement, the number of constraint limitations in residential energy management problems depends on the controlled device number.

\textbullet \textit{ Integrated electricity-gas energy system:}
The integrated EMSs can combine electricity, heat, cooling, natural gas, and hydrogen to achieve the efficient synergy of various carriers for meeting energy demands, based on conversion, distribution, and storage technologies. Generally, controllable devices in integrated EMSs are divided into three types: 1) RENs, e.g., wind generator (WG) and solar photovoltaic (PV); 2) storage systems, e.g., hydrogen storage system (HSS) and thermal energy storage (TES); 3) energy conversion devices, e.g., combined heat and power (CHP), electric heat pump (EHP), and gas boiler (GB). Hence, the action space can be designed as:
\begin{align}
    &a^\text{Int} =[\alpha^\text{WG},\alpha^\text{PV},\alpha^\text{CHP},\alpha^\text{EHP},\alpha^\text{GB},\beta^\text{HSS},\beta^\text{TES}]^\intercal,
\end{align}
where $\alpha^\text{WG},\alpha^\text{PV},\alpha^\text{CHP},\alpha^\text{EHP},\alpha^\text{GB}\in[0,1]$ represent the magnitude of output heat power of WGs, PVs, CHPs, EHPs, and GBs, as a percentage of their maximum power limits. Variables $\beta^\text{HSS},\beta^\text{TES}\in[-1,1]$ represent the charging/discharging power rate (positive/negative) of HSS and TES as a percentage of their power capacities. The designed action space is continuous because all the above devices can be controlled continuously. Further, the state space can be defined to capture device operating information as:
\begin{align}
    &s^\text{Int} = [t, E^\text{HSS},E^\text{TES},P^\text{WG},P^\text{PV},\boldsymbol{\lambda}^\text{Int}, \textbf{D}^\text{Int}]^\intercal,
\end{align}
where $E^\text{HSS},E^\text{TES}$ are the measured state-of-charge of HSS and TES, reflecting the environment dynamics after the action; $P^\text{WG}$ and $P^\text{PV}$ are the maximal WG and PV generation power determined by stochastic weather conditions; $\boldsymbol{\lambda}^\text{Int}$ denotes the pre-offered grid prices for various energy types, such as electricity/gas/carbon prices; $\textbf{D}^\text{Int}$ represents the uncertain energy demands for various energy types, e.g., electricity and heat demands. If more uncertainties are considered in the integrated network, e.g., real-time price, it becomes more challenging to solve the designed MDP.

For the action space in integrated EMSs, decisions for each device are independent without correlation, which may lead to violations of energy balance constraints. Hence, apart from the regular upper/lower limits for single variables, e.g., import/export power capacity, the inner demand-supply balance of heat/electricity is also a significant constraint, which can be generalized as:
\begin{align}
    & H^\text{d} = H^\text{s}, P^\text{d} = P^\text{s}, G^\text{d} = G^\text{s},
\end{align}
where $H^\text{d}, P^\text{d}, G^\text{d}$ are heat, power, and gas demands; $H^\text{s}, P^\text{s}, G^\text{s}$ denote corresponding heat, power, and gas supplies in integrated EMSs. Note that the generalized formulation only shows the design principle, and specific constraint models should be specified based on the inner structure of integrated energy systems.

\textbullet \textit{ Grid-level power dispatch:}
Due to the security and economy, the optimal power flow (OPF), is the fundamental tool underlying extensive scenarios of grid-level power dispatch, especially security-constrained OPF\cite{yi2023real}. The key control objects in OPF are the power outputs of generators, thus the action space for grid-level power dispatch is generally defined as:
\begin{align}
    & a^\text{Opf} = [P^\text{gen}_k, Q^\text{gen}_k | k\in\mathcal{N}_\text{G}]^\intercal,
\end{align}
where $P^\text{gen}_k$ and $Q^\text{gen}_k$ represent the commands to the active and reactive power outputs of $k$-th generator. Commonly, the on/off statuses of the generators are assumed to be predetermined and not changed during real-time dispatch. The on/off statuses can also be included as a binary decision in action space\cite{fan2022soft}.
The state space is usually defined as:
\begin{align}
    & s^\text{Opf} = [P^\text{gen}_k, Q^\text{gen}_k, v_i, P^\text{D}_i, Q^\text{D}_i, \tilde{P}^\text{pre}, \tilde{Q}^\text{pre}| k\in\mathcal{N}_\text{G}, n\in\mathcal{N}_\text{B}]^\intercal,
\end{align}
where $P^\text{D}_i$ and $Q^\text{D}_n$ are the active and reactive net demand at bus $i$; $\tilde{P}^\text{pre}$ and $\tilde{Q}^\text{pre}$ are the prediction of the next active and reactive system loads. Decision-making in OPF is highly related to the accuracy of future load forecasting. To improve accuracy, some works tend to increase the length of the forecasting time slots in the state, providing more historical information \cite{yi2023real}. Nevertheless, a large scale of the state space increases the training complexity, which requires a trade-off between information content and complexity.

The constraints considered in OPF problems mainly include the power flow equations, bus voltage/transmission line flow limits, and physical limits of controllable generators, which contain equality and inequality constraints. Hence, the constraint can be generalized as:
\begin{subequations}
\label{OPF_constraint}
    \begin{align}
        &P^\text{G}_i-P^\text{D}_i = v_i\sum\limits_{j\in\mathcal{N}_\text{B}}v_j(G_{ij}\cos\theta_{ij}+B_{ij}\sin\theta_{ij}),\forall i\in\mathcal{N}_\text{B},\\
        &Q^\text{G}_i-Q^\text{D}_i = v_i\sum\limits_{j\in\mathcal{N}_\text{B}}v_j(G_{ij}\sin\theta_{ij}-B_{ij}\cos\theta_{ij}),\forall i\in\mathcal{N}_\text{B},\\
        &\underline{P}^\text{gen}\leq P^\text{gen}_k\leq \overline{P}^\text{gen}_k, 
        \underline{Q}^\text{gen}\leq Q^\text{gen}_k\leq \overline{Q}^\text{gen}_k,\quad \forall k\in\mathcal{N}_\text{G},\\
        &\underline{v}\leq v_i\leq \overline{v},\quad \forall i\in\mathcal{N}_\text{B},\label{voltage_opf_constraint}
    \end{align}
\end{subequations}
where $P^\text{G}_i$ and $Q^\text{G}_i$ are the injections of the active and reactive power at bus $i$; $G_{ij}$ and $B_{ij}$ are the conductance and susceptance of the transmission line between bus $i$ and bus $j$; $\theta_{ij}$ is the angle difference between bus $i$ and bus $j$; $\overline{P}^\text{gen}_k$/$\underline{P}^\text{gen}$ and $\overline{Q}^\text{gen}_k$/$\underline{Q}^\text{gen}$  are the corresponding upper/lower bounds of the active and reactive power outputs of $k$-th generator.

\subsubsection{Reward and cost functions}
The control objective of OPF problems is to minimize the total generation costs of the power system, where the reward function for cost minimization can be generally written as:
\begin{align}
    &r^\text{Em} = - \sum\nolimits_{t=0}^\infty\sum\nolimits_{k\in\mathcal{N}_\text{G}}
    \left[ a_k(P^\text{gen}_k)^2 + b_k P^\text{gen}_k +c_k \right],
\end{align}
where $a_k$, $b_k$, and $c_k$ are the operation cost coefficients of generation $k$. Considering that Eq. (\ref{OPF_constraint}) includes lots of operation constraints, the cost function is usually designed as the total penalization for violations, which can be generalized as:
\begin{align}
    &c^\text{Em} = \sum\nolimits_{|l|}\left[ w_l\cdot \text{ReLU}(c_l - \overline{c}_l) \right],
\end{align}
where $l$ represents the index of the constraints defined in Eq. (\ref{OPF_constraint}), i.e., $|l|=3\mathcal{N}_\text{B}+2\mathcal{N}_\text{G}$; $w_l$ denotes the penalty weight of each constraint; function $\text{ReLU}(x)=\max (0,x)$ is a linear rectification function for violation measurements; $c_l$ and $\overline{c}_l$ represent the actual power flow state and required limit value. Taking constraint Eq. (\ref{voltage_opf_constraint}) as an example, the $c_l$ and $\overline{c}_l$ can be defined by $c_l=\{v_i, -v_i\}$ and $\overline{c}_l=\{\overline{v}, -\underline{v}\}$.

\subsubsection{Safe RL techniques}
Energy management covers multiple energy flows among different subjects. Based on different problem characteristics, we present the applied safe RL techniques in EMSs, to show the advantages of various techniques under different scenarios.

\textbullet \textit{ Safe layer:}
The model-based safe layer can effectively cope with hard constraints in EMSs. For instance, reference \cite{ceusters2023adaptive} proposed \textit{OptLayerPolicy} to increase the accuracy of the cost function, which can keep a high sample efficiency at the initial stage to solve the closest feasible action in safe layers. Besides, based on a specific and well-designed correction rule, authors in \cite{yu2023district} also successfully address the hard constraints through linear programming relying on the local constraint model.
Some recent works propose model-free safe layers to release the model requirement. In \cite{wang2023secure}, because it is hard to assess the operation safety of a multi-energy microgrid without models, authors learn a dynamic security assessment by NNs to abstract a physical-informed safety layer on top of the conventional RL framework, which manner can maintain the secure operation of physical constraints. NNs are effective approaches and are commonly used in EMSs to assess the safety of complex systems, including the device-level control in residential energy management\cite{huang2022mixed}.
In addition, a hybrid safety layer that combines model knowledge and data-driven methods is proposed in \cite{yi2023real} to solve security-constrained OPF, where the projection model is initialized with prior knowledge and updated continuously based on collected data. This hybrid safe layer can enhance system security in the early learning stage, compared with the model-free safe layer.

\textbullet \textit{ Policy optimization:}
Most safe RL techniques applied in EMSs, by changing policy optimization criteria, are model-free because they involve the large-scale gradient calculation for policy networks. For instance, Li et al. adopted the CPO algorithm in \cite{li2021online} and \cite{li2019constrained} to find a safety-guaranteed scheduling policy for microgrid energy management and EV scheduling strategy, respectively. Reference \cite{sayed2023online} proposed model-free CSAC for integrated EMSs to find optimal energy flow in real-time operation, which is safe and with less hyperparameter sensitivity. In addition, the Lagrange method can also be applied to handling soft physical constraints in integrated EMSs. Authors in \cite{hu2024multi} combined the idea of primal-dual and Lagrange multipliers to solve the high-dimensional non-convex problem, for the operational optimization of integrated community energy systems. 
A few studies have tried to introduce model-based policy gradients for safe policy optimization, for safety-critical problems. For example, in a security-constrained OPF problem, reference \cite{yan2022hybrid} proposed a hybrid data-driven method to approximate policy gradients by solving Karush–Kuhn–Tucker conditions. For optimal power management of networked microgrids, authors in \cite{zhang2020multi} derived a safe policy gradient based on the AC power flow equations, to transform the non-convex problem into a tractable convex iterative quadratically constrained linear program.

\subsubsection{Discussion}\leavevmode

\textbullet \textit{ Insufficient scenario occurrence:}
Some energy management problems for power systems, such as service restoration, probably have insufficient online training due to the infrequent occurrence (e.g., low outage rates). Hence, the requirement for a large amount of training scenarios is one of the major impediments to applying safe RL, especially for safety-critic but infrequent scenarios. Pre-training can be considered a compensatory for safe RL, as an alternative to the agent's early training process. To this end, reference \cite{du2022deep} proposed a two-stage safe RL framework by introducing the pre-training stage before the online training stage, where the initialized agent can gain a jump-start performance through expert imitation. More online approaches for handling scenario insufficiency issues remain to be explored.

\textbullet \textit{ Multi-agent safe control for multi-EMSs:}
Energy management problems usually contain multiple devices or distributed microgrids for cooperation control, involving global and local constraints. Most aforementioned studies consider all the constraints in one cost function and address them through one RL agent. Reference \cite{zhang2020multi} proposed a multi-agent consensus-based training algorithm for distributed microgrids, which designs multiple autonomous controllers for joint safe control through local communication. To further decouple intractable power-carbon flow constraints for low-carbon EMSs, reference \cite{zhang2024networked} extended the traditional CPO algorithm to a consensus multi-agent CPO, to achieve the safe control for low-carbon demand management.

\textbullet \textit{ Spatial-temporal perception:}
Accurate perception of the spatial-temporal operating characteristics is significant for EMSs, such as the spatial distribution of power flows and the temporal evolution of the renewables. For the energy management in distributed microgrids, reference \cite{ye2023safe} proposed an interior-point policy optimization (IPO) algorithm to utilize a logarithmic CBF to ensure constraint safety, by introducing edge-conditioned convolutional and long short-term memory networks. These two feature extraction networks can effectively find out the spatial and temporal dependencies, for better state prediction accuracy.

\subsection{Other applications}
In addition to the above three critical applications, the safe RL concept is also applied to other problems in power systems, including dynamic distribution network reconfiguration\cite{gao2020batch}, real-time congestion management\cite{yang2023dynamic}, electricity market\cite{rokhforoz2023safe}, setpoint optimization for transmission systems\cite{tarle2023safe}, organic Rankine cycle system control\cite{zhang2022event}, resilient proactive scheduling\cite{liang2021safe}, transmission overload relief\cite{cui2023online}, emergency recovery under line outages\cite{weiss2023safe}, etc..

\section{Challenges and perspectives}\label{cases}
Safe RL, as one type of RL variant, naturally faces all challenges in traditional RL, such as data availability and scalability summarised in \cite{Lina_RL_review}. In this section, we do not repeat the common challenges in traditional RL, and only present the unique challenges of safe RL applications in power systems, including four aspects: (1)convergence and optimality; (2) training efficiency; (3) universality; and (4) real-world environment deployment. Three future directions are then discussed.

\subsection{Convergence and optimality}
The RL applications aim to find an optimal control policy after training, thereby ensuring the successful convergence of training and policy optimality are two key criteria to evaluate the reliability and effectiveness of the algorithm.
For the training convergence in safe RL, introducing safe layers or changing policy update rules both limit the RL agent's exploration space, and even probably bring wrong feedback, which results in the traditional RL convergence theory no longer being applicable. 
For instance, the RL agent's original action is usually corrected in safe layers, which rely on specific design principles for different tasks. Hence, it is impossible to prove that a random safe layer design can bring a successful convergence. This is because a too strong intervention in a safe layer can easily destroy the convergence of training, if the reward feedback is not properly corrected by the safe layer.
Currently, some researchers have presented approximate proof of convergence or the theoretical guarantee under certain conditions, e.g., Lyapunov-based and CBF-based safe layer methods\cite{chow2019lyapunov, CBF_safe_layer}. However, these proofs are usually proposed based on certain assumptions, such as an approximation of the environment model, a limitation of the policy space, or a simplification of the optimization process. 
Therefore, we believe that rigorous global convergence proof remains a challenge for safe RL, and future research needs to explore more general theoretical frameworks to provide stronger guarantees of training convergence.

For the policy optimality in safe RL, because RL training processes update policy mainly relying on the feedback of random explorations, there exists a conflict and trade-off between the conservative policy for safety and aggressive explorations for improvement. This trade-off may result in a poor reward expectation in some scenarios, obtaining a sub-optimal or even bad policy. Specifically, a safe but limited policy-searching area makes the agent unable to explore the entire state/action spaces, which probably causes the algorithm to converge to a locally optimal solution rather than a global one. This phenomenon is similar to the fundamental dilemma between exploration and exploitation in RL. Considering the research on safe RL algorithms is still in its early stage, few papers are handling this challenge effectively through algorithm design or optimization strategy. 

Besides, the actual power system control involves not only a single object but requires cooperation between several distributed areas. Purely single-agent safe RL may be too hard to handle all global and local constraints and suffer from convergence issues. Currently, there are very few solutions that offer effective learning algorithms for safe multi-agent control problems. Recently, reference\cite{kuba2021trust} proposed the first multi-agent trust region method that successfully attains theoretical guarantees of both reward improvement and satisfaction of safety constraints. Then, as the first safety-aware model-free algorithms, reference \cite{gu2023safe} extend CPO and Lagrange methods to the multi-agent area with theoretical analysis. Despite limited theoretical work on this subject, investigating multi-agent safe RL for multi-device control on cooperative tasks in power systems is envisioned to be an important future direction.

\subsection{Training efficiency}
The training efficiency for safe RL mainly includes two types: sample efficiency and computation efficiency, where sample efficiency refers to the utilization of data samples (i.e., high sample efficiency can obtain a better policy through fewer data samples), and computation efficiency indicates the utilization of computing resources (e.g., CPU, GPU, and memory).
Compared with conventional RL, sample efficiency in safe RL is decreased significantly when facing safety-critical constraints, because variance among collected samples becomes smaller when explorations are limited within a safe area. That is, a smaller variance provides less knowledge for the RL agent and slows down its policy update iteration. 
Moreover, in practice, power systems are considered reward-sparse environments for safety-critical problems, because unsafe data samples are usually not adequate or even limited, such as voltage/frequency violations. The reward-sparse issue also influences the sample efficiency.
Currently, authors in \cite{chen2021safe} proposed a sample-efficient safe RL framework to achieve efficient learning with limited samples through three techniques: 1) avoiding behaving overly conservatively;  2) encouraging safe exploration;  3) treating RL agents as expert demonstrations.

In addition, the introduced safe module increases the computation complexity, which further decreases the computation efficiency of safe RL algorithms. 
For instance, when using MPC-based methods, one needs to solve an extra linear or quadratic program at every time step $t$\cite{MPC_safe_layer}. When using CBF-based methods, one needs to store lots of previous RL policy networks and solve multiple separate quadratic programs in sequence to evaluate each CBF controller\cite{CBF_safe_layer}. 
When requiring real-time operations in power systems, computation efficiency becomes an issue for the safe RL application, especially in highly non-linear and complex environments.

\subsection{Universality}
The power system is a large-scale dynamic system whose system characteristics may change according to dynamic demands, such as the changing number of generators/transformers, and the variational topology brought by line faults/maintenance. In particular, the increasing distributed energy resources may more frequently change the grid topology for distribution networks. Hence, the universality of a well-trained agent is necessary for power grids to cope with various operation scenarios. However, most of the current safe RL algorithms have poor universality, leading to low reusability of controllers. Once the power system model changes, it may be necessary to retrain the agent, resulting in expensive computation and training costs. The key reasons for poor universality in different safe RL techniques are not the same.

\textit{ 1) Universality for safe layers:}
For safe layer-based techniques, all constraints are designed based on an assumed accurate or approximated model, which will directly influence the solved optimization problem results in safe layers at each time step. When the system dynamic changes, the constraint formulation is necessary to be re-derived to fit the new system for safety. Thus, the intervention method for the agent by safe layers is changed, and the agent should be re-trained based on the designed new safe layers.

\textit{ 2) Universality for changing optimization criteria:}
Although this category of safe RL techniques (introduced in Section \ref{sec:2.3}) is a totally model-free method without prior model knowledge, its training relies on the collected historical costs for constraint violations. When the system dynamic changes, the relationship between the real-time cost and the system operation state is also changed correspondingly, which is different from the original data distribution. Thus, the original safe policy is probably not safe for the new system, requiring re-training based on newly collected data.

Considering that RL is fundamental and vibrant research that garners significant attention, innovations in RL are emerging at a rapid pace, e.g., inverse RL \cite{tang2022multi}, meta RL \cite{yu2020meta}, hierarchical RL \cite{jendoubi2023multi}, attention-based RL \cite{xie2023multi}, etc.. 
To enhance the algorithm universality, one can fully leverage existing research achievements in reinforcement learning. For instance, hierarchical RL can decompose the control task into multiple subtasks to design dynamic safety constraints and exploration methods, reducing the probability of the policy falling into local optima. Meta RL can be used to collect safety information from multiple different environments, thereby enhancing safety performance in downstream task training without sufficient expert experience.

\subsection{Real-world deployment}
For most published works, case studies are designed in simulated environments based only on small-scale grids with a few control objects. To the best of our knowledge, safe RL has not yet been implemented in any real-world power system, even though it claims to be safer than traditional RL algorithms. 
One of the crucial limitations in safe RL deployment is the initial policy quality, since an initialized policy without training cannot ensure performance even though it may be safe. Since the control in grids usually involves multi-process coupling and multi-department cooperation, it is difficult to directly deploy a random and poor initial policy in the real world for such a long training time. One potential direction to improve the reliability of initial policies is to obtain a pre-trained policy in simulators first to find satisfactory initial policies \cite{lesage2022batch}. 
However, because of the lack of theoretical assurance, this idea also faces safety issues caused by the gap between simulators and real-world large-scale systems. Two state-of-the-art techniques can be combined with safe RL frameworks to further address the reality gap: sim2real\cite{yavas2023real} and digital twin modeling\cite{xu2023delay}. Specifically, sim2real is a method for transferring safe RL algorithms trained in simulated environments to real-world applications, and digital twin modeling is a technique for creating a virtual copy of a real-world physical system in a digital space.
For instance, simulations in reference \cite{liu2023effective} are implemented in the digital twin environment to present effective energy management for household demand response, employing safe RL and fuzzy reasoning. The proposed digital twin model presents a real-time consumer interface, including smart devices, energy price signals, smart meters, solar PVs, batteries, electric vehicles, and grid supply.

In addition, exploring the effective integration of online safe RL techniques with the deployment of offline RL presents a promising research direction. This idea is conducted solely on pre-collected offline trajectory datasets, without the need for real-time interaction with the environment. While this approach circumvents unsafe exploration phases, ensuring safety during model training, there still exists a gap between the offline training data and real-world datasets. To date, only the constraints penalized Q-learning (CPQ) method \cite{xu2022constraints} has proposed using additional cost critic (such as reward critic) to learn constraint values, which effectively bridges the distribution gap between the offline and real-world datasets. However, CPQ still has a theoretical error bound under mild assumptions, as an offline method.

\section{Conclusions}\label{conclude}
This paper provides a comprehensive review of safe RL techniques and applications in power systems for the first time. We summarize two categories of state-of-the-art safe RL techniques, which are based on safe layers and policy optimization criteria. Then, three key applications in power systems are summarized through the detailed design of state, action, reward, cost, and applied safe RL methods. Finally, several key challenges and future directions are discussed.

In summary, although safe RL has been paid more attention to for better application in power systems, there is still quite a long distance from real-world deployment. The most important issue is the lack of theoretical proof for safe RL applications, which cannot ensure the safety and optimality of common scenarios. In fact, as an online training-based control method, explorations cannot be completely avoided in safe RL techniques, which makes it hard to deploy individually for safety-critical scenarios. Considering the advantage of model-free characteristics, combining safe RL with model-based traditional controllers is probably more promising and practical.

\bibliographystyle{ieeetr}
\bibliography{ref}

\end{document}